\documentclass{article}

\usepackage{graphics}
\usepackage{amsmath}
\usepackage{amsfonts}
\usepackage{psfrag}
\usepackage{graphicx,epsfig,float}

\newenvironment{pf}
{\noindent\textit{\bf{Proof}}\newline}{\vspace{10pt}}
\def\qq#1{\hskip #1 em \relax}

\def\qed{\vrule height7.5pt width4.17pt depth0pt}
\newcommand{\smallo}{{\scriptstyle\cal O}}
\newcommand{\Rb}{\mathbb{R}}

\newcommand{\Nb}{\mathbb{N}}

\newcommand{\beqa}{\begin{eqnarray}}
\newcommand{\eeqa}{\end{eqnarray}}
\newcommand{\beq}{\begin{equation}}
\newcommand{\eeq}{\end{equation}}

\newcommand{\xri}{x \rightarrow \infty}
\newcommand{\trmi}{t \rightarrow +\infty}
\newcommand{\srmi}{s \rightarrow +\infty}
\newcommand{\taurmi}{\tau \rightarrow +\infty}
\newcommand{\jri}{j \rightarrow \infty}
\newcommand{\simf}{\varphi_3(\tau+\xi\sqrt{\tau}, \tau)}
\newcommand{\Mp}{M_{\psi}}
\newcommand{\nn}{\nonumber}
\newcommand{\ve}{\varepsilon}
\newcommand{\vf}{\varphi}

\newcommand{\ds}{\displaystyle}
\newcommand{\BD}{Becker-D\"{o}ring}

\newcommand{\const}{\operatorname{{constant}}} 
\newcommand{\sgn}{\operatorname{{sgn}}} 

\let\tilde=\widetilde

\begin{document}
\newtheorem{defn}{Definition}
\newtheorem{cor}{Corollary} 
\newtheorem{lem}{Lemma}
\newtheorem{conj}{Conjecture}
\newtheorem{prop}{Proposition}
\newtheorem{theo}{Theorem}

\title{Long-Time Behaviour and Self-Similarity in a 
Coagulation Equation With Input of Monomers
\thanks{Discussions with W. Oliva, J.T. Pinto, and C. Rocha, are 
greatfully acknowledged. FPdC was partially supported by 
projects FCT MAT/199/94 20199-POCTI/FEDER, and CRUP Ac\c{c}\~ao Luso-Brit\^anica B-14/02. 
HvR was supported by NSERC of Canada. JADW is grateful to the British Council Treaty of Windsor
Programme 2002/3}
}

\author{F.P. da Costa\\ 
Universidade Aberta, DCET\\
Rua Fern\~ao Lopes 9, 2\raisebox{.7ex}{{\small \underline{o}}}Dto, P-1000-132 Lisboa, Portugal\\
{\small and}\\
Instituto Superior T\'ecnico, CAMGSD\\ Av. Rovisco Pais 1, 
P-1049-001 Lisboa, Portugal\\
\\
H.J. van Roessel\\
University of Alberta, Dep. of Mathematical and Statistical Sciences\\ Edmonton, Canada T6G 2G1\\ \\
J.A.D. Wattis\\ University of Nottingham, School of Mathematical Sciences\\
University Park, Nottingham NG7 2RD, UK}

\date{{\small \text{{\sc to appear in:}}\hspace*{1mm} \text{Markov Processes and Related Fields}\\
\text{{\sc submited:}}\hspace*{1mm} \text{February 1, 2005}\\ 
\text{{\sc revised:}}\hspace*{1mm} \text{August 30, 2005}
}}

\maketitle

\begin{abstract}
For a coagulation equation with \BD\/ type interactions and time-independent 
monomer input we study the detailed long-time behaviour of nonnegative 
solutions and prove the convergence to a self-similar function. 

\vspace*{.5cm}\noindent

\vspace*{.5cm}\noindent

\end{abstract}

\flushbottom
\newpage

\section{Introduction} 

A number of differential equations providing mean-field descriptions for the kinetics
of particle aggregation have been the focus of much mathematical effort
(see \cite{LM} for a review of recent mathematical developments.) An important 
special case of this type of equations are the well known coagulation equations 
first studied by the Polish physicist Marian Smoluchowski (1872-1917)
\cite{smol}: denoting
by $c_j=c_j(t)$ the concentration at time $t$ of an agglomerate of $j$ identical 
particles, and assuming binary aggregation following mass action law is the only
process taking place, we obtain the discrete version of Smoluchowski's system
\beq
\dot{c}_j =\frac{1}{2}\sum_{k=1}^{j-1}a_{j-k,k}c_{j-k}c_{k}-
c_j\sum_{k=1}^{\infty}a_{j,k}c_k
\label{Smol}
\eeq
where the first sum in the right-hand side is defined to be zero when $j=1.$
The kinetic coefficients $a_{j,k}$ measure the efficiency of the reaction between
$j-$clusters and $k-$clusters to produce $(j+k)$-clusters, and, as such, they should 
satisfy the minimal requirements of symmetry and nonnegativity $a_{j,k}=a_{k,j}\geq 0.$
The precise form of the coefficients depends on the physical phenomena being
modelled (see, eg \cite[Table 1]{dc98}.)

A central problem in the study of the long-time behaviour of solutions to
(\ref{Smol}) is their convergence to some similarity profile $\psi,$
when $j$ and $t\rightarrow +\infty,$
$$c_j(t) \sim s(t)^{-\beta}\psi(j s(t)^{-\alpha})$$
where $\alpha$ and $\beta$ are appropriate positive exponents, and $s(t)$ a scaling 
function. 

This is conjectured to occur, for a large class of initial conditions,
in the case of homogeneous kinetic coefficients, i.e., those satisfying
$a_{j,k}= K(j,k)$, with kernels $K$ for which
$K(ux, uy)=u^{\lambda}K(x, y).$
For a very recent survey see \cite{Ley03}.

From a rigorous point of view not much is known, although some
results were previously available \cite{dC96,kp}, and a number of significant advances
have very recently been made \cite{emrr,FL04,mp03-1,mp03-2}.

In the present paper we are interested in studying this type of behaviour for
a class of non-homogeneous coefficients that are also relevant in the applications: 
the Becker-D\"{o}ring type coefficients satisfying $a_{j,k}=0$ if $\min\{j, k\}>1.$
With this type of restriction system (\ref{Smol}) is sometimes called the
{\it addition model\/} \cite{BK,HE,L99}, and is written in the form 
\beq
\left\{
\begin{array}{lcl}
\dot{c}_1 & = & - a_1c_1^2 - c_1 \displaystyle{\sum_{j=1}^{\infty}a_jc_j}\\
\dot{c}_j & = & a_{j-1}c_1c_{j-1}-a_{j}c_1c_j, \quad j \geq 2 
\end{array}
\right.\label{Addi}
\eeq
where $a_1=\frac{1}{2}a_{1,1},$ and $a_j=a_{j,1}$ if $j\geq 2.$
It is a special case of the Becker-D\"{o}ring coagulation equations \cite{bcp} 
(without fragmentation of clusters.) Physically this corresponds to cases
where the only effective reactions are those involving at least a $1-$particle cluster
(a {\it monomer\/}.) The addition model is not expected to support self-similar behaviour 
in the sense outlined above because the special
role played by the monomers implies the dynamics gets frozen when one
runs out of monomers, and the limit state will not be related to any universal
profile (however see \cite{BK}.) Thus a prerequisite for a meaningful study of similarity
behaviour in the addition model is the consideration of some mechanism
that constantly provides new monomers to the system. The most widely studied
of these mechanisms is fragmentation \cite{bcp} in which case
difficult problems concerning the large time evolution of large clusters, not
dissimilar from the dynamic scaling behaviour refered to above, have
only recently started to be tackled rigorously \cite{LM02,N03}. Another mechanism,
is to externally provide for the increase of monomers by adding to the
right-hand side of the monomer equation in (\ref{Addi}) a source term $J_0(t).$ A case
like this was actually what the original system proposed by Becker and D\"oring
amounted to \cite{bd,p89}, since they considered the case where the concentration of monomers
stay constant in time. Physically this can only be implemented by coupling
the system to a monomer bath of infinite size. Another way to externally supply
monomers, which is much more reasonable in a number of pratical applications, such
as the modelling of polymerisation processes, is to {\it a priori\/} give the source 
term $J_0(t)$, independently of the existing cluster concentrations. 
The simplest such situation is when $J_0(t) = \const$, and this has indeed been 
used in applications, for instance in mean-field models for epitaxial thin 
layer growth \cite{b94,be96,e96,grc}.

The study of these Becker-D\"oring, or Smoluchowski, systems with input of
monomers, with or without fragmentation, has greatly progressed in the 
mathematical modelling literature (see, for example, \cite{lk,pa,w04,wbc}.)
A very recent work by one of us \cite{w04} provides a fairly extensive
study in the case of constant coagulation and fragmentation coefficients
and monomer input rates given by $J_0(t)=\alpha t^{\omega}.$ A variety of
possible similarity profiles was observed, depending on the balance between
coagulation and fragmentation and also on the rate of monomer input, $\omega.$
Our present paper has a much more limited goal: we intend to rigorously
analyse the constant coagulation $(a_j=1)$ addition model with constant
monomer input $(\omega=0)$, namely
\beq
\left\{
\begin{array}{lcl}
\dot{c}_1 & = & \alpha - c_1^2 - c_1 \displaystyle{\sum_{j=1}^{\infty}c_j}\\
\dot{c}_j & = & c_1c_{j-1}-c_1c_j, \quad j \geq 2 
\end{array}
\right.\label{1}
\eeq
where $\alpha>0$ is independent of $t.$ 

This is a very special case of the systems studied by \cite{w04}, but it is,
not only sufficiently simple to allow a rigorous mathematical analysis
of the similarity behaviour to be performed, even including an higher
order analysis, but still is of some interest for applications \cite{be96}.
Anyway, it should just be considered as a first step towards a
rigorous mathematical analysis of the behaviour uncovered in the 
mathematical modelling literature cited above.

A point that is a clear consequence of our analysis is the way
the dynamical behaviour of solutions to this infinite dimensional system
is really determined by a very small dimensional quantity, 
namely by the dynamics in a unidimensional centre manifold.  This
reduction of the determining modes of the system is the mathematical
counterpart to the loss of memory of the initial data implied by the physical
assumption of convergence to self-similarity. It is reasonable to believe that 
this will also be the case in more complicated systems, but actually, at present,
this geometric approach to self similarity in coagulation problems is not
yet entirely clear.

We now describe the contents of the paper, and briefly discuss their import.

Our main goal is to understand the large time behaviour of solutions to (\ref{1}), in
particular the precise rates of convergence and possible existence of self-similarity.
For a solution to exist we require that the sum appearing in (\ref{1}) be
convergent. In particular, it need not have finite mass $M_1(t):=\sum_{j=1}^{\infty}jc_j(t).$
If we introduce the total number of clusters as a new macroscopic
variable $c_0(t)$ defined by $$ c_0(t)=\sum_{j=1}^{\infty}c_j(t),$$ and
formally differentiate termwise, we conclude that $c_0$ satisfies the evolution
equation $\dot{c}_0=\alpha-c_0c_1$.  Thus, system (\ref{1}) can, at least
formally, be written, in closed form, as
\beq
\left\{
\begin{array}{lcl}
\dot{c}_0 & = & \alpha - c_0 c_1,\\
\dot{c}_1 & = & \alpha - c_1^2 -c_0 c_1,\\
\dot{c}_j & = & c_1c_{j-1}-c_1c_j, \quad j \geq 2.
\end{array}
\right.\label{2}
\eeq

In the next section we show that our formal calculations are justified and that
the solutions of system (\ref{2}) are in fact solutions to the original system (\ref{1}). 
This is done by the use of a generating function. 

The study of the long time behaviour of solutions requires the knowledge of the behaviour of 
$c_1$ and this can be obtained through a
detailed analysis of the two-dimensional system for the
monomer concentration and total number of clusters. This is done in Section~\ref{sec3}
and involves the use of invariant regions and of a technique related to the Poincar\'e 
compactification method followed by the application of centre manifold theory. 
The analysis of the full system (\ref{2}) is made possible by the fact that an
appropriate change of time $t\mapsto \tau$ changes the $c_j$ equation to the linear
differential equation $\tilde{c}_j\,' = \tilde{c}_{j-1}-\tilde{c}_{j}$ from where
we easily get a representation formula for $\tilde{c}_j$ in terms of the non monomeric
initial data and of $\tilde{c}_1$ (see expression (\ref{12}) in Section~\ref{sec:equiv}.)
It is this fortunate  fact, together with the information on $c_1$ proved in
Section~\ref{sec3}, that allow us to study the long time behaviour of the
other components $c_j(t)$ establishing that all decay as $t^{-1/3}$ when $t\rightarrow\infty,$ 
for all $j\geq 1,$ which will be proved in Section~\ref{sec4}. 
The same approach, but involving a higher degree of technical difficulties, is
used to prove the convergence of solutions to a
self-similar profile, namely $\tilde{c}_j(\tau) \sim \tau^{-1/2}\Phi_1(j/\tau),$ and the precise 
determination of the function $\Phi_1.$ This is valid for all solutions with
initial data bounded above by $\rho j^{-\mu}$ with $\mu>1/2,$ which clearly
includes all the physically interesting finite mass initial data cases. 
The precise statement and proof takes the whole of Section~\ref{sec5}.
The main technique involved is a controlled asymptotic evaluation of the
sum and the integral appearing in the representation formula for $\tilde{c}_j.$
It turns out that the similarity profile $\Phi_1(\eta)$ has a discontinuity
at $\eta=1$ (see Figure 1.)

%
%
%
%
%
%
\begin{figure}[h]\label{FiguraIntro1}
\begin{center}
\psfrag{eta}{$\eta$}
\psfrag{geta}{$\Phi_1(\eta)$}
\includegraphics[scale=0.7]{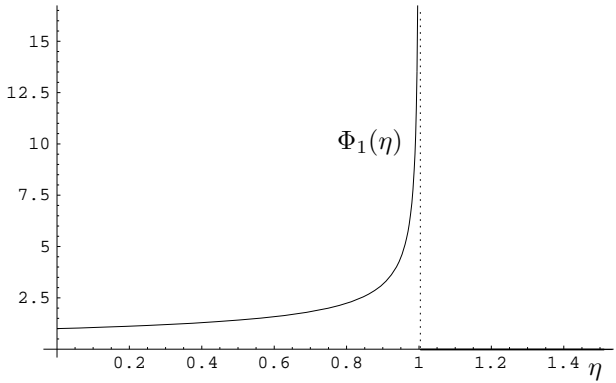}
\end{center}
\caption{Graph of the similarity profile $\Phi_1(\eta)$.}
\end{figure}
%
%
%
%
%
%

To help understand why this is so, it is perhaps interesting to call the reader's attention 
to an heuristic, non-rigorous, but nevertheless enlightening, geometric picture
of what is happening when we look for similarity behaviour in our system. Considering the 
$\tilde{c}_j$ system for very large $j$ and assuming our level of description
is such that we can consider $j$ as a continuous positive real variable, the equation
for $\tilde{c}$ becomes the conservation law
\beq
\partial_{\tau}\tilde{c}+\partial_j\tilde{c}=0\qquad (j, \tau)\in\Rb^+\times\Rb^+. \label{cl}
\eeq
Propagation of initial and boundary data for (\ref{cl}) occurs along the characteristic 
direction $j=\tau.$ This implies that initial data for (\ref{cl}), defined in $\Rb^+\times\{0\}$,
is not at all felt in the region of $\Rb^+\times\Rb^+$ with $\eta:=j/\tau <1,$ and reciprocally,
boundary data for (\ref{cl}), given in $\{0\}\times\Rb^+$, is not felt where $\eta>1.$ 
Since, in the present case, the search for self-similar behaviour entails us looking
for limits of the solution along lines of constant slope $j/\tau$, the geometric picture
provided by the conservation law leads us to anticipate that the limits of our
representation formula for $\tilde{c}_j,$ taken with $\eta<1$ will demand 
less stringent conditions on the initial data than limits with $\eta>1,$ and
reciprocally for the requirements on the ``boundary data'' $\tilde{c}_1(\tau).$
And of course the border case $\eta=1$ is expected to exhibit some problems.
This turns out to be exactly what happens.

Naturally, the discontinuity in $\Phi_1$ is an indication that along the characteristic
direction $\eta=1$ the solutions do not scale as $\tau^{-1/2}$ (if they scale at all.)
In fact, we end the paper by proving, in Section~\ref{sec6}, that there is another
similarity variable, $\xi:=\frac{j-\tau}{\sqrt{\tau}},$
that essentially corresponds to a kind of inner expansion of the characteristic 
direction $\eta=1,$ and a similarity profile, $\Phi_2(\xi),$ such that, 
at least for monomeric initial data, solutions behave like
$\tilde{c}_j(\tau) \sim \tau^{-1/4}\Phi_2(\tfrac{j-\tau}{\sqrt{\tau}}),$ where the function  $\Phi_2$ (to be
defined in Section~\ref{sec6}) can be expressed \cite{w04} in terms of Kummer's 
hypergeometric functions \cite[pp 504-515]{AS}, and have the graph presented in Figure 2.
It is worth noting the similitude of the graphs of the scaling functions in
Figures 1 and 2 to those in \cite[Fig. 2]{be96} and \cite[Fig. 2, 8]{e96}.

%
%
%
%
%
%
\begin{figure}[ht]\label{FiguraIntro2}
\begin{center}
\psfrag{xi}{$\xi$}
\psfrag{fxi}{$\Phi_2(\xi)$}
\includegraphics[scale=0.7]{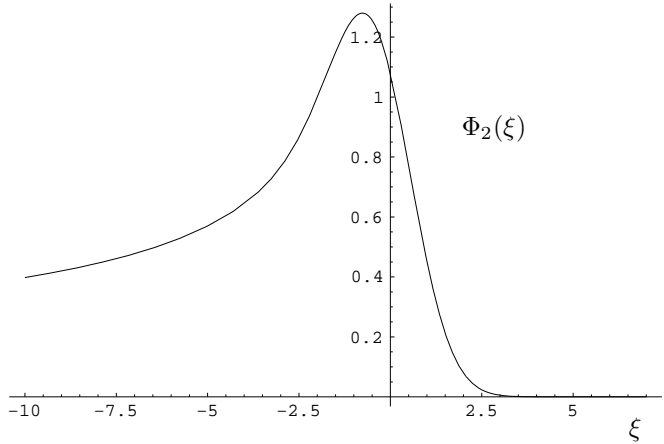}
\end{center}
\caption{Graph of the similarity profile $\Phi_2(\xi)$}
\end{figure}
%
%
%
%
%
%

The restriction to monomeric initial data is probably not important, but we
could so far not overcome the difficulties involved in controlling the initial data term
for {\it small\/} cluster sizes. We discuss these difficulties at the end of
Section~\ref{sec6}. 

Putting together the results in the last two Sections of this paper,
we can conclude that solutions do converge to a similarity profile,
that is a unimodal distribution with a spike located at $j \sim t^{2/3}$
decaying to zero like $t^{-1/3}$ for $j$s behind the spike, and the spike itself
decaying like $t^{-1/6}.$ This means that the family of solutions $\{c_j(t)\}$ tends
to become more spiky as we look for larger and larger times $t$ and cluster sizes $j.$

\section{Equivalence of the related system}\label{sec:equiv}   

Let $(c_j(t))_{j=0}^\infty$ be a nonnegative solution of (\ref{2}).  To
show that this is also a nonnegative solution of the original
addition system (\ref{1}) requires that we prove the
convergence of $\sum_{n=1}^{\infty}{c}_n(t)$ to ${c}_0(t)$.
To this end, it is convenient to introduce a new time scale
\begin{equation}\label{eq:tau}
\tau(t):=\int_0^t\!c_1(s)\,ds
\end{equation}
along with scaled variables
\begin{equation}\label{eq:ctilde}
\tilde{c}_j(\tau):=c_j(t(\tau)),
\end{equation}
where $t(\tau)$ is the inverse function of $\tau(t)$. Since $c_1(t)>0,$ these 
are well defined and $\tau$ is an increasing function of $t.$
Now consider the $c_j$-equations in (\ref{2}):
$$
\dot{c}_j = c_1c_{j-1}-c_1c_j, \qquad j\geq 2.
$$
By the change of variables $(t, c_j(t))\mapsto (\tau, \tilde{c}_j(\tau))$ these
equations can be written as a linear system
$$
\tilde{c}_j\mbox{}' = \tilde{c}_{j-1}-\tilde{c}_j, \qquad j\geq 2,
$$
where $(\cdot)' = \frac{d}{d\tau}$. Now this system of ordinary differential equations 
is a lower triangular system that can be solved recursively starting with the $j=2$ equation.
Doing this we obtain, by the variation of constants formula,
\begin{equation}\label{12}  
\tilde{c}_j(\tau) = e^{-\tau}\sum_{k=2}^{j}\frac{\tau^{j-k}}{(j-k)!}c_k(0) +
\frac{1}{(j-2)!}\int_0^{\tau}\tilde{c}_1(\tau-s)s^{j-2}e^{-s}ds.
\end{equation}
We now prove the following.

\begin{theo}\label{th:exist}
If the series of initial data is convergent, ie if
$\sum_{j=1}^{\infty}c_j(0)<\infty$, then a solution of  system (\ref{2}) will
also be a solution of the original system (\ref{1}).
\end{theo}
\begin{pf}
Introduce the following generating function:
\begin{equation}\label{eq:Fdef}
F(\tau,z) := \sum_{n=1}^{\infty}\tilde{c}_n(\tau)z^n.
\end{equation}
Using (\ref{12}) we can rewrite this as follows:
\begin{equation*}
F(\tau,z) = \tilde{c}_1(\tau)z + \sum_{n=2}^{\infty}\tilde{c}_n(\tau)z^n 
= \tilde{c}_1(\tau)z + G(\tau,z) + H(\tau,z),
\end{equation*}
where
\begin{align*}
G(\tau,z) &:= e^{-\tau}\sum_{n=2}^{\infty}\sum_{k=2}^{n}\frac{\tau^{n-k}z^n}{(n-k)!}c_k(0), \\
H(\tau,z) &:= \sum_{n=2}^{\infty}\frac{z^n}{(n-2)!}\int_{0}^{\tau}\tilde{c}_1(\tau-s)s^{n-2}e^{-s}\,ds.
\end{align*}
Examining these expressions, we find that the series can actually be summed.  For $G$ we have
\begin{align*}
G(\tau,z) &= e^{-\tau}\sum_{n=2}^{\infty}\sum_{k=0}^{n-2}\frac{\tau^{n-k-2}z^n}{(n-k-2)!}c_{k+2}(0)
=  e^{-\tau}\sum_{k=0}^{\infty}\sum_{m=0}^{\infty}\frac{\tau^{m}z^{m+k+2}}{m!}c_{k+2}(0) \\
&=  e^{-\tau}\sum_{k=0}^{\infty}\left(\sum_{m=0}^{\infty}\frac{(\tau z)^{m}}{m!}\right)z^{k+2}c_{k+2}(0)
= e^{-\tau(1-z)}\sum_{k=2}^{\infty}z^{k}c_{k}(0).
\end{align*}
By hypothesis the series in the above expression converges for $|z|\le1$, so we have
\begin{equation}\label{eq:G}
G(\tau,z) = e^{-\tau(1-z)}\left(F(0,z)-\tilde{c}_1(0)z\right) \qq2\textrm{for }|z|\le1.
\end{equation}
As for $H$ we have
\begin{align}\label{eq:H}
H(\tau,z) &= z^2\int_{0}^{\tau}\tilde{c}_1(\tau-s)\left(
        \sum_{n=2}^{\infty}\frac{(sz)^{n-2}}{(n-2)!}\right)e^{-s}\,ds\notag \\
&= z^2\int_{0}^{\tau}\tilde{c}_1(\tau-s)e^{sz} e^{-s}\,ds
= z^2 \int_{0}^{\tau}\tilde{c}_1(s)e^{-(\tau-s)(1-z)}\,ds.
\end{align}
The expression for $F$ now becomes:
\begin{equation}\label{eq:F}
F(\tau,z) = \tilde{c}_1(\tau)z + e^{-\tau(1-z)}\left(F(0,z)-\tilde{c}_1(0)z\right)
        + z^2 \int_{0}^{\tau}\tilde{c}_1(s)e^{-(\tau-s)(1-z)}\,ds,
\end{equation}
which, at $z=1$, yields
\begin{equation}\label{eq:c0}
F(\tau,1) = \tilde{c}_1(\tau) + F(0,1)-\tilde{c}_1(0) + \int_{0}^{\tau}\tilde{c}_1(s)\,ds.
\end{equation}
One can easily verify that $F(\tau,1)$ given in (\ref{eq:c0}) satisfies the
first equation of (\ref{2}) which proves that $F(\tau,1)=\tilde{c}_0(\tau)$.  \hfill \qed
\end{pf}

\vspace{10pt}
Having established that solutions of (\ref{2}) are in fact solutions of
(\ref{1}), all subsequent analysis will be carried out on system (\ref{2}).

\section{The bidimensional ODE system governing the monomer dynamics}   
\label{sec3}

{}From the reduced system (\ref{2}) we observe that the equations governing both the monomer 
dynamics and the total number of clusters are actually the bidimensional $(c_0, c_1)$-system, 
the dynamics of which can be studied quite independently of the remaining components.
To simplify notation we will use $x=c_1$ and $y=c_0$ in the study of this bidimensional
system.

Let $\alpha>0$ be a constant, and consider the system
\beq
\left\{
\begin{array}{lcl}
\dot{y} & = & \alpha - xy\\
\dot{x} & = & \alpha - x^2-xy
\end{array}
\right.\label{3}
\eeq
We are interested in nonnegative solutions to (\ref{3}) and so, from hereon, everytime
we speak of solutions we actually mean nonnegative solutions. Our first result concerns
the gross features of the long time behaviour of solutions. Another result, to
be presented in Proposition~\ref{prop2}, will establish some finer details of the long time behaviour.

\begin{prop}\label{prop1}
For any solution $(x, y)$ of (\ref{3}) the following holds true as $\trmi$:
$x(t)\rightarrow 0,\; y(t)\rightarrow +\infty,$ and $x(t)y(t)\rightarrow \alpha$.
\end{prop}

\noindent
{\bf{Proof}}: Let $\Omega$ be the connected subset of $\Rb^+\times\Rb^+$ whose boundary
is $\{y=0\}\cup\{x=0\}\cup\{xy=\alpha\}.$ Consider first our initial data in 
$\overline{\Omega}$. Elementary phase plane analysis shows this set is positively
invariant for the flow (see Figure 3.) 

%
%
%
%
%
%
\begin{figure}[ht]
\begin{center}
\includegraphics[scale=0.2]{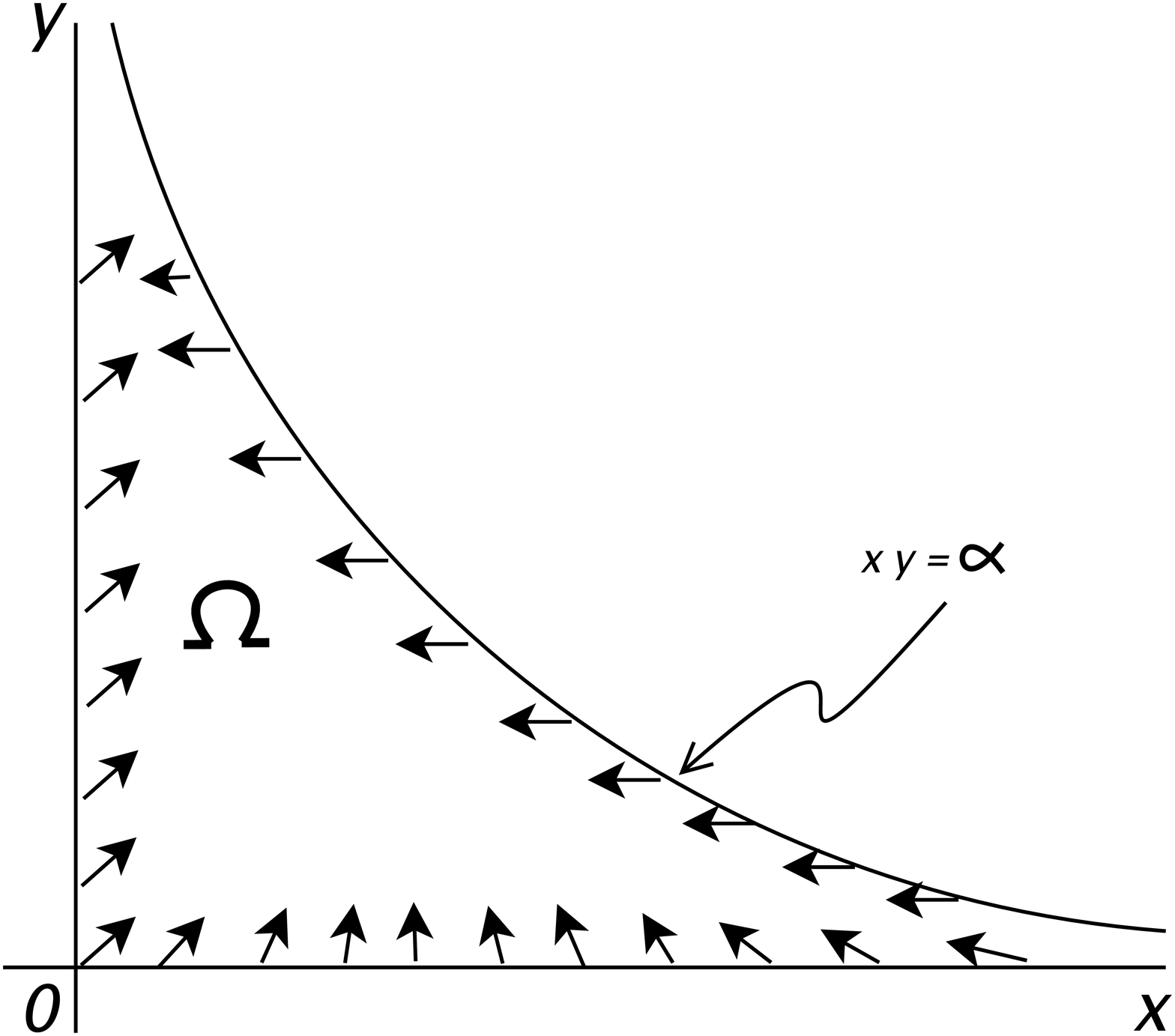}
\end{center}
\caption{Region $\Omega$ with a sketch of the flow in $\partial \Omega.$}
\end{figure}
%
%
%
%
%
%

We now observe that (\ref{3})   
does not have equilibria, and consider the subset
$\Omega_1$ of $\Omega$ defined by
$$\Omega_1 := \left\{(x, y)\in \Omega: \max\left\{0, \frac{\alpha}{x}-x\right\}
\leq y \leq \frac{\alpha}{x}\right\}.$$
Again, elementary phase plane analysis shows that $\Omega_1$ is positively invariant
and, for any initial condition in $\Omega$ the corresponding orbit will
eventually enter $\Omega_1$ (see Figure 4.)

%
%
%
%
%
%
\begin{figure}[ht]
\begin{center}
\includegraphics[scale=0.2]{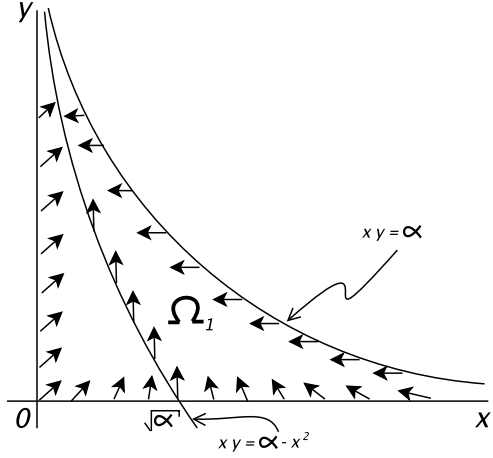}
\end{center}
\caption{Region $\Omega_1$ with a sketch of the flow in $\partial \Omega,
\partial\Omega_1$ and in $\Omega.$}
\end{figure}
%
%
%
%
%
%

{}From this we immediately conclude that, as $\trmi$, we have 
$x(t)\rightarrow 0$
and $y(t)\rightarrow +\infty$. Furthermore, for all initial data in $\Omega,$
there exists a $T$ (depending on the initial condition) such that, for all
$t>T$, the orbit is in $\Omega_1,$ and so
$$t>T \Rightarrow \frac{\alpha}{x}-x\leq y\leq \frac{\alpha}{x} \Leftrightarrow 
\alpha-x^2\leq  xy \leq \alpha.$$
Since we know that $x(t)\rightarrow 0$ as $\trmi$, applying limits in the above 
inequality gives
$$\lim_{\trmi}x(t)y(t)=\alpha.$$

Consider now initial data $(x_0, y_0)\in \Omega_2=\Rb^+\times\Rb^+\setminus\Omega.$
Fix $K_1>x_0$, $K_2>y_0$ and let $\Omega_2(K_1, K_2)=\Omega_2\cap ([0, K_1]
\times[0, K_2]).$ By the analysis of the flow in $\partial\Omega_2(K_1, K_2)$ we
conclude that the orbit will eventually enter $\Omega_1$ (see Figure 5) and so
the previous analysis apply. 

%
%
%
%
%
%
\begin{figure}[ht]
\begin{center}
\includegraphics[scale=0.2]{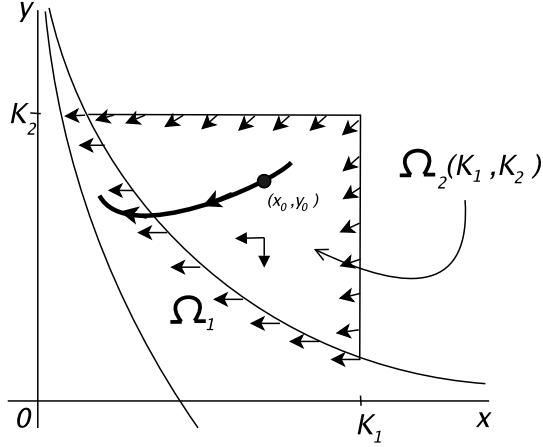}
\end{center}
\caption{Flow in the set $\Omega_2(K_1, K_2)$ surrounding an initial 
point $(x_0, y_0)$ outside $\Omega.$}
\end{figure}
%
%
%
%
%
%

This concludes the proof. \hfill\qed
\vspace*{4mm}

In the next Proposition we prove results about the rate
of convergence of $x(t)$ and $y(t)$ as $\trmi,$ using an
approach akin to Poincar\'e compactification.

\begin{prop}\label{prop2}
For any solution $(x, y)$ of (\ref{3}) we have:
\begin{description}
\item[{\tt (i)}] $\ds{\lim_{\trmi}\left(\frac{3}{\alpha}t\right)^{1/3}\!\!\!x(t)=1}$
\item[{\tt (ii)}] $\ds{\lim_{\trmi}\left(3\alpha^2t\right)^{-1/3}y(t)=1}$
\item[{\tt (iii)}] $\ds{\lim_{\trmi}\left(\frac{3}{\alpha}t\right)^{2/3}\!\!\!(\alpha-x(t)y(t))=1}$
\end{description}
\end{prop}

\noindent
{\bf{Proof}}: For the present proof we found convenient to work with the variable
$v:=\alpha-xy$ instead of $y$, and with a new time scale. The idea is essentially that used
in Poincar\'e compactification: to bring to the origin the critical point at infinity, and
to turn it into a point of the phase plane by appropriately changing the time scale
(cf. eg \cite[chap. 5]{z92}.) By changing variables $(x, y)\mapsto (x, v)$ system 
(\ref{3}) becomes
\beq
\left\{
\begin{array}{lcl}
\dot{x} & = & v-x^2\\
\dot{v} & = & \alpha x -2 xv-\alpha\frac{v}{x}+\frac{v^2}{x}.
\end{array}
\right.\label{4}
\eeq
Suppose $x(0)\neq 0$. Otherwise, since we know by the phase plane analysis in the
proof of Proposition~\ref{prop1} that $x(t)>0$ for all $t>0,$ just redefine time so
that $x(0)$ becomes positive.
Now change the time scale
$$
t\mapsto \zeta = \zeta(t):= \int_0^t\frac{1}{x(s)}ds,
$$
and define $\left(\tilde{x}(\zeta), \tilde{v}(\zeta)\right):= \left(x(t(\zeta)), v(t(\zeta))\right),$
where $t(\zeta)$ is the inverse function of $\zeta(t)$. By Proposition~\ref{prop1},
we have $x(t)\rightarrow 0$ as $\trmi$, and so also $\zeta\rightarrow +\infty$ as $\trmi.$ 
With the new time scale system (\ref{4}) becomes
\beq
\left[
\begin{array}{l}
\tilde{x} \\ \tilde{v}
\end{array}
\right]'  = \left[ \begin{array}{cr} 0 & 0 \\ 0 & -\alpha\end{array} \right]
\left[\begin{array}{c}\tilde{x}\\ \tilde{v}\end{array}\right] + 
\left[\begin{array}{c}\tilde{v}\tilde{x}-\tilde{x}^3 \\ \alpha \tilde{x}^2
+\tilde{v}^2 -2\tilde{v}\tilde{x}^2\end{array}\right]
\label{4'}
\eeq
where  $(\cdot)' = \frac{d}{d\zeta}.$ The region of interest, corresponding to
$(x, y)\in \Rb^+\times\Rb^+,$ is $(x, v)\in \Rb^+\times (-\infty, \alpha),$
but actually (\ref{4'}) is valid in all of the phase plane $\Rb^2.$ In this way, the
critical point at infinity of system (\ref{3}) is mapped to the critical point at the
origin for (\ref{4'}). From the results in Proposition~\ref{prop1} we know
that all orbits of (\ref{4'}) obtained from orbits of (\ref{3}) by
the above map will eventually enter  $\Rb^+\times (0, \alpha)$ for sufficiently
large times $\zeta$ and converge to $(0, 0)$ as $\zeta\rightarrow +\infty.$
Using standard results \cite[chap. 2]{c81} it is straightforward to conclude the
existence of a centre manifold for (\ref{4'}) given by
\beq
\tilde{v}= \phi(\tilde{x})\quad {\rm where} \quad \phi(\tilde{x}) = \tilde{x}^2 -
\frac{1}{\alpha}\tilde{x}^4 + {\cal O}(\tilde{x}^6),\label{centremanifold}
\eeq
that locally attracts all orbits. The dynamics on the centre manifold
is given by 
$$
\tilde{x}_c' = -\frac{1}{\alpha}\tilde{x}_c^5 + {\cal O}(\tilde{x}_c^7)\quad{\rm as}\;\;\zeta\rightarrow
+\infty.
$$
Write this differential equation as
\beq
\frac{d}{d\zeta}\left(\frac{\alpha}{4\tilde{x}^4_c}\right)= 1+{\cal O}(\tilde{x}_c^2)\quad 
{\rm as}\;\;\zeta\rightarrow +\infty.\label{5'}
\eeq
Fix an initial condition $(\zeta_0, \tilde{x}_0)$ and fix $\varepsilon >0$ arbitrarily.
Since we know that $\tilde{x}_c(\zeta)\rightarrow 0$ as $\zeta\to+\infty$ we conclude that
there exists a $T>\zeta_0$ such that, for all $\zeta>T,$ the following  inequalities hold
$$
1-\varepsilon \leq \frac{d}{d\zeta}\left(\frac{\alpha}{4\tilde{x}^4_c}\right) \leq 1+\varepsilon.
$$
Integrating these differential inequalities between $T$ and $\zeta$ we get
\beq
\frac{\alpha}{4\tilde{x}^4_T}+(1-\varepsilon)(\zeta-T)\leq \frac{\alpha}{4\tilde{x}^4} \leq
\frac{\alpha}{4\tilde{x}^4_T}+(1+\varepsilon)(\zeta-T),\label{5''}
\eeq
where $\tilde{x}_T =\tilde{x}(T).$
Dividing (\ref{5''}) by $\zeta$ and taking  $\ds{\varliminf_{\zeta\to+\infty}}$
and $\ds{\varlimsup_{\zeta\to+\infty}}$ we obtain
$$
1-\varepsilon\leq \varliminf_{\zeta\to+\infty}\frac{\alpha}{4\zeta \tilde{x}_c^4}\leq 
 \varlimsup_{\zeta\to+\infty}\frac{\alpha}{4\zeta \tilde{x}_c^4}\leq 1+\varepsilon
$$
and since $\varepsilon>0$   is arbitrary, we conclude that
\beq
\lim_{\zeta\to+\infty}\left(\frac{4}{\alpha}\zeta\right)^{1/4}\!\!\tilde{x}_c(\zeta)=1 \label{xclim}
\eeq
for solutions corresponding to orbits of (\ref{4'}) {\it on\/}
its centre  manifold. 
{}From standard centre manifold theory \cite[chap. 2]{c81}, long time behaviour of 
$(\tilde{x}(\zeta), \tilde{v}(\zeta))$ is determined
by the behaviour on the centre manifold modulo exponentially decaying terms
${\cal O}\left(e^{-\lambda \zeta}\right),$ where $\lambda\in (0, \alpha)$,
in particular we can write
$$\tilde{x}(\zeta) = \tilde{x}_c(\zeta) + {\cal O}\left(e^{-\lambda \zeta}\right).$$
Multiplying this equality by  $\left(\frac{4}{\alpha}\zeta\right)^{1/4}$ and using
(\ref{xclim}) we conclude that
\beq
\lim_{\zeta\to+\infty}\left(\frac{4}{\alpha}\zeta\right)^{1/4}\!\!\tilde{x}(\zeta)=1.\label{xlim}
\eeq
We can do the same for the evolution of the $\tilde{v}$ variable: for orbits on the
centre manifold we have  $\tilde{v}_c = \tilde{x}_c^2 + {\cal O}(\tilde{x}_c^4),$
multiplying this by $\left(\frac{4}{\alpha}\zeta\right)^{1/2},$ and using (\ref{xclim})
we conclude that 
$$
\lim_{\zeta\to+\infty}\left(\frac{4}{\alpha}\zeta\right)^{1/2}\!\!\tilde{v}_c(\zeta)=1. 
$$
To obtain the dynamics outside the centre manifold we proceed as with the
$\tilde{x}$ variable and get
\beq
\lim_{\zeta\to+\infty}\left(\frac{4}{\alpha}\zeta\right)^{1/2}\!\!\tilde{v}(\zeta)=1. \label{vlim}
\eeq
In order to obtain the corresponding rate estimates in the original time variable $t$
we need to relate the asymptotics of both time scales. By the definition of the new time scale
$$\zeta(t) -\zeta(t_0) =  \int_{t_0}^t\!\frac{1}{x(s)}ds$$
and so $\frac{d\zeta}{dt}= \frac{1}{x(t)}.$ Hence, by the inverse function theorem,
 $\frac{dt}{d\zeta}= \tilde{x}(\zeta),$ and thus
\beq
t(\zeta)-t(\zeta_0) = \int_{\zeta_0}^{\zeta}\tilde{x}(s)ds \label{tofzeta}
\eeq
where $\zeta_0=\zeta(t_0).$ From (\ref{xlim}) we know that
\beq
\forall\ve>0,\,\exists T=T(\ve):\,\forall \zeta >T,\, 
\left(\frac{4}{\alpha}\zeta\right)^{1/4}\!\!\tilde{x}(\zeta)\in [1-\ve, 1+\ve]
\label{bound}
\eeq
Let us look first at the upper bound: for all $\zeta >T,$
$$
\tilde{x}(\zeta) \leq \left(\frac{4}{\alpha}\zeta\right)^{-1/4}(1+\ve),
$$
substituting this into (\ref{tofzeta}) we get, for $\zeta_0\geq T,$
\beqa
t(\zeta)-t(\zeta_0) & \leq & \int_{\zeta_0}^{\zeta}
\left(\frac{4}{\alpha}s\right)^{-1/4}\!\!(1+\ve)ds \nn \\
& = & (1+\ve)\left(\frac{\alpha}{4}\right)^{1/4}\frac{4}{3}\left(\zeta^{3/4}-\zeta_0^{3/4}\right).\nn
\eeqa
Multiplying this inequality by $\frac{3}{4}\left(\frac{4}{\alpha}\right)^{1/4}\zeta^{-3/4}$
and passing to the limit as $\zeta\to+\infty$ results in
$$
\varlimsup_{\zeta\to+\infty}\frac{3}{4}\left(\frac{4}{\alpha}\right)^{1/4}\!\!\zeta^{-3/4}t(\zeta) \leq 1+\ve.
$$
Using the lower bound of (\ref{bound}) and the same argument we obtain the reverse bound
$$
\varliminf_{\zeta\to+\infty}\frac{3}{4}\left(\frac{4}{\alpha}\right)^{1/4}\!\!\zeta^{-3/4}t(\zeta) 
\geq 1-\ve.
$$
and by the arbitrariness of $\ve>0$ we get
\beq
\frac{3}{4}\left(\frac{4}{\alpha}\right)^{1/4}\!\!\zeta^{-3/4}t(\zeta) = 1+\smallo(1)\quad\text{as $\zeta\to+\infty$},
\label{zetat}
\eeq
which, together with (\ref{xlim}) and (\ref{vlim}),  allow us to conclude that, as 
$\trmi$ (ie, as $\zeta\to+\infty$)
$$
\left(\frac{3}{\alpha}t\right)^{1/3}\!\!\!x(t) =\left(\frac{4}{\alpha}\zeta\right)^{1/4}\!\!\!
\tilde{x}(\zeta)(1+\smallo(1))
= 1+\smallo(1), 
$$
and
$$
\left(\frac{3}{\alpha}t\right)^{2/3}\!\!\!v(t) =\left(\frac{4}{\alpha}\zeta\right)^{1/2}\!\!\!
\tilde{v}(\zeta)(1+\smallo(1))
= 1+\smallo(1), 
$$
thus proving {\tt (i)} and {\tt (iii)} respectively.

Finally, to prove {\tt (ii)}, observe that for the original variable $y(t)$ we have 
$y=\frac{\alpha-v}{x}$ and thus, as $\trmi$,
$$
(3\alpha^2t)^{-1/3}y(t) = \frac{1}{\left(\frac{3}{\alpha}t\right)^{1/3}\!\!x(t)}-
\frac{\left(\frac{3}{\alpha}t\right)^{2/3}\!\!v(t)}{\left(\frac{3}{\alpha}t\right)^{1/3}\!\!x(t)}\,
\frac{1}{\alpha\left(\frac{3}{\alpha}t\right)^{2/3}}\rightarrow 1
$$
which concludes the proof.\hfill\qed
\vspace*{4mm}

\section{Long time behaviour of the system}   
\label{sec4}

We now turn our attention to the long-time behaviour of $c_j(t)$ with
$j\geq 2$.  It is convenient to consider the time scale introduced in Section
\ref{sec:equiv}.  The first result we need is the following,
the proof of which is entirely analogous to that of the corresponding results
presented in the second half of the proof of Proposition~\ref{prop2}, and so we
refrain from repeating it here,

\begin{prop}\label{prop:ctauasymp}
With $(c_j)$, $\tau$, and $(\tilde{c}_j)$ as given in (\ref{eq:tau}) and
(\ref{eq:ctilde}), the following holds true:
\begin{description}
\item[{\tt (i)}] $\ds{\lim_{\trmi}\frac{2}{3}\left(\frac{3}{\alpha}\right)^{1/3}\!\!\!t^{-2/3}\tau(t)=1}$;
\item[{\tt (ii)}] $\ds{\lim_{\taurmi}\sqrt{\frac{2\tau}{\alpha}\,}\;\tilde{c}_1(\tau)=1}$;
\end{description}
\end{prop}

With this knowledge, we can now use (\ref{12}) to get information on the long time behaviour of 
$\tilde{c}_j(\tau)$ for $j\geq 2,$ specifically, we prove now that {\tt (ii)} in Proposition~\ref{prop:ctauasymp} holds for all $\tilde{c}_j(\tau)$. To see this, multiply (\ref{12}) by
$\sqrt{\frac{2\tau}{\alpha}}.$ The first term in the right hand side of the
resulting expression is the contribution due to the non monomeric initial data, and since $j$ 
is fixed, we have, as $\taurmi$,
\[
\tau^{\frac{1}{2}}e^{-\tau}\sum_{k=2}^{j}\frac{\tau^{j-k}}{(j-k)!}c_k(0) = {\cal O}\left(\tau^{j-\frac{3}{2}}e^{-\tau}\right) = \smallo\left(e^{-\lambda\tau}\right),
\]
for every $\lambda<1.$ In order to study the remaining integral term, first change the integration 
variable $s\mapsto y=s/\tau$ in order to get the integral in a fixed bounded region, and
define the function $\psi(\cdot) :=\sqrt{\frac{2}{\alpha}\,(\cdot)\,}\,\tilde{c}_1(\cdot).$
We thus get
\[
\sqrt{\frac{2\tau}{\alpha}}\frac{1}{(j-2)!}\int_0^{\tau}\tilde{c}_1(\tau-s)s^{j-2}e^{-s}ds
= \frac{\tau^{j-1}}{(j-2)!}\int_0^{1}\frac{\psi(\tau(1-y))y^{j-2}}{\sqrt{1-y}}e^{-\tau y}dy.
\]
Let $0<\ve<1$ be fixed, and write the integral as $\int_0^{1-\ve}+\int_{1-\ve}^{1}.$ The last
integral is fairly easy to handle: 
Since $\psi(s)$ is a continuous function and, by
Proposition~\ref{prop:ctauasymp}, is $1+\smallo(1)$ as $\srmi,$ we conclude that it is bounded 
in $[0, \infty)$ and so there exists a positive constant $\Mp$ such that $0\leq \psi(s)\leq \Mp$ for
all $s.$ From this it follows that
\beqa
\tau^{j-1}\int_{1-\ve}^{1}\frac{\psi(\tau(1-y))y^{j-2}}{\sqrt{1-y}}e^{-\tau y}dy & \leq &
\Mp\tau^{j-1}e^{-(1-\ve)\tau}\int_{1-\ve}^{1}\frac{y^{j-2}}{\sqrt{1-y}}dy  \nonumber \\
& < & \Mp\tau^{j-1}e^{-(1-\ve)\tau}\int_{1-\ve}^{1}\frac{1}{\sqrt{1-y}}dy  \nonumber \\
& < & 2\sqrt{\ve}\Mp\tau^{j-1}e^{-(1-\ve)\tau}, \nonumber
\eeqa
and so it is also exponentially small when $\taurmi.$ For the integral over $(0, 1-\ve)$ we use
$y<1-\ve \Rightarrow \tau(1-y)>\tau\ve\rightarrow +\infty$ as $\taurmi,$ and 
Proposition~\ref{prop:ctauasymp} {\tt (ii)} to conclude that $\psi = 1+\smallo(1)$ in the
region of integration, provided $\tau$ is sufficiently large. Therefore, 
$\forall\delta>0,\,\exists T(\delta):\,\forall \tau>T(\delta),\, 
\psi(\tau(1-y)) \in [1-\delta, 1+\delta]$, and, as $\taurmi,$
\beq
(1-\delta)I_{0,j}(\tau) \leq 
\int_0^{1-\ve}\frac{\psi(\tau(1-y))y^{j-2}}{\sqrt{1-y}}e^{-\tau y}dy
\leq (1+\delta)I_{0,j}(\tau) \label{antewatson}
\eeq
with
\[
I_{0,j}(\tau):= \int_0^{1-\ve}\frac{y^{j-2}}{\sqrt{1-y}}e^{-\tau y}dy.
\]
This integral is easily estimated using Watson's Lemma \cite[pp 427-8]{af03}: from the 
Taylor expansion 
\[
\frac{y^{j-2}}{\sqrt{1-y}} = y^{j-2}\sum_{k=0}^{\infty}\frac{(2k-1)!!}{(2k)!!}y^k,
\]
which converges for $|y|<1,$ we can apply Watson's Lemma directly to get
\[
I_{0,j}(\tau) = \frac{\Gamma(j-1)}{\tau^{j-1}} + {\cal O}\left(\tau^{-j}\right),\quad\mbox{\rm as}\;\taurmi.
\]
Plugging this into (\ref{antewatson}) immediately results in 
\[
\frac{\tau^{j-1}}{(j-2)!}\int_0^{1-\ve}\frac{\psi(\tau(1-y))y^{j-2}}{\sqrt{1-y}}e^{-\tau y}dy
=1 + {\cal O}\left(\tau^{-1}\right)\quad\mbox{\rm as}\;\taurmi.
\]
This, together with the other two exponentially small contributions proved above,
implies that $\sqrt{\frac{2\tau}{\alpha}\,}\;\tilde{c}_j(\tau)\longrightarrow 1,$ as
$\taurmi,$ for all $j\geq 2,$ thus complementing part {\tt (ii)} of 
Proposition~\ref{prop:ctauasymp}. Using part {\tt (i)} we can easily translate this behaviour
to the original time scale, and conclude the following:

\begin{theo}
Let $(c_j)$ be any non-negative solution of (\ref{1}) with initial data
satisfying $c_0(0)=\sum_{j=1}^{\infty}c_j(0)<\infty$. Then,
as $\trmi$, we have
\begin{description}
\item[{\tt (i)}] $\ds{\left(\frac{3}{\alpha}t\right)^{1/3}\!\!\!c_j(t)\longrightarrow 1}$ \qq1 
                                                                            for all $j\ge1$;
\item[{\tt (ii)}] $\ds{\left(3\alpha^2t\right)^{-1/3}\sum_{j=1}^{\infty}c_j(t)\longrightarrow 1}$;
\item[{\tt (iii)}] $\ds{\left(\frac{3}{\alpha}t\right)^{2/3}\!\!
        \left(\alpha-c_1(t)\sum_{j=1}^{\infty}c_j(t)\right)\longrightarrow 1}$.
\end{description}
\end{theo}

\section{Self-similar behaviour of the coagulation system outside the characteristic direction}   
\label{sec5}

We have now reached the main objective of the paper: the study of convergence to
similarity profiles in this addition model.

Let $\Phi_1: \Rb^+\setminus\{1\}\rightarrow \Rb$ be the function given by
$$
\Phi_1(\eta):=\begin{cases} \ds{\frac{1}{\sqrt{1-\eta}}}& \text{if $\eta<1$},\\
0 & \text{if $\eta>1$}.\end{cases}
$$
The main result in this part is the following:


\begin{theo}\label{mainteo1}
Let $(c_j)$ be any non-negative solution of (\ref{1}) with initial data satisfying
$\exists \rho>0, \mu>\frac{1}{2}: \forall j, c_j(0)\leq\rho/j^{\mu}$.
Let $\tau(t)$ and $\tilde{c}_j(\tau)$ be as given in (\ref{eq:tau}) and (\ref{eq:ctilde}) respectively.
Then,
$$
\lim_{\substack{j,\,\tau \rightarrow +\infty\\ \eta=j/\tau\; {\rm fixed}\\ \eta \neq 1}}
\sqrt{\frac{2}{\alpha}\tau\,}\;
\tilde{c}_j(\tau) = \Phi_1(\eta)
$$
\end{theo}


\noindent
{\bf{Proof}}\\
Expression (\ref{12}) describes the time evolution of $\tilde{c}_j(\tau)$ by the sum
of a contribution dependent only on the non-monomeric part of the  initial data with a term
determined by the behaviour of $c_1(t)$ in the appropriate time scale. For monomeric initial
data, only this last term is relevant, and we start our analysis by it.

\subsection{Monomeric initial data} 

Assuming monomeric initial conditions we have, for $j\geq 2,$ 
\beq
\sqrt{\frac{2}{\alpha}\tau\,}\;\tilde{c}_j(\tau)
 = \frac{\sqrt{\frac{2}{\alpha}\tau\,}}{(j-2)!}
\int_0^{\tau}\tilde{c}_1(\tau-s)s^{j-2}e^{-s}ds.\label{13}
\eeq
Consider the function $\vf_1$ defined in $[2, \infty)\times[0,\infty)$ by
\beq
\vf_1(x, \tau) := \frac{\sqrt{\frac{2}{\alpha}\tau\,}}{\Gamma(x-1)}
\int_0^{\tau}\tilde{c}_1(\tau-s)s^{x-2}e^{-s}ds.\label{14}
\eeq
Clearly $\vf_1 = \sqrt{\frac{2}{\alpha}\tau}\,\tilde{c}\;$ if $2\leq x=j\in \Nb$, and the use of $\vf_1$
is a good deal more convenient in order to obtain the required asymptotic results. 
So we now concentrate on (\ref{14}). Let $\eta:=x/\tau.$ 
By changing variable $s\mapsto y=s/\tau,$ using 
the recursive relation $\Gamma(x-1)=\Gamma(x)/(x-1),$ and Stirling's
asymptotic formula $\Gamma(x)=e^{-x}x^{x-\frac{1}{2}}\sqrt{2\pi}\left(1+{\cal O}
(x^{-1})\right)$ as $\xri$,
 we can write, as $\taurmi,$ 
\beq
\vf_1(\eta\tau, \tau) = \frac{\eta^{\frac{3}{2}-\eta\tau}\tau^{\frac{1}{2}}}{\sqrt{\alpha\pi}}
\left(1+{\cal O}\left(\tau^{-1}\right)\right)
\int_0^{1}\frac{\tilde{c}_1(\tau(1-y))\tau^{1/2}e^{\tau(\eta\log y-y+\eta)}}{y^2}dy.\label{15}
\eeq
In order to make use of our knowledge about the long time behaviour of $\tilde{c}_1(\tau)$
given in Proposition 3 it is convenient to rearrange the integrand of (\ref{15}) by multiplying
and dividing it by $\sqrt{\frac{2}{\alpha}(1-y)}$, resulting in the following, as $\taurmi,$
\beq
\vf_1(\eta\tau, \tau) = \frac{1}{\sqrt{2\pi}}\eta^{\frac{3}{2}-\eta\tau}\tau^{\frac{1}{2}}
\left(1+{\cal O}\left(\tau^{-1}\right)\right)
\int_0^{1}\psi(\tau(1-y))\frac{e^{\tau(\eta\log y-y+\eta)}}{y^2\sqrt{1-y}}dy,\label{16}
\eeq
where $\psi(\cdot) :=\sqrt{\frac{2}{\alpha}\,(\cdot)\,}\,\tilde{c}_1(\cdot)$ is the function 
that was already defined and used in Section~\ref{sec4}. 
The proof now reduces to the asymptotic evaluation of the function
\beq
I(\eta, \tau) := \tau^{\frac{1}{2}}\eta^{-\eta\tau}e^{\tau\eta}\int_0^{1}\psi(\tau(1-y))\frac{e^{\tau(\eta\log y-y)}}{y^2\sqrt{1-y}}dy
\label{I}
\eeq
as $\taurmi.$

We first consider the case $\eta>1.$

Observe that $y^{-2}e^{\tau(\eta\log y-y)} =e^{(\eta\tau-2)\log y-\tau y} =: e^{g_1(y)},$ 
where $g_1:\Rb^+\rightarrow\Rb$, defined by the last equality, satisfies, for all $y\in(0,1]$ and 
$\tau>\frac{2}{\eta-1}$
$$
g_1'(y) = \frac{\eta\tau-2}{y}-\tau \geq (\tau\eta-2)-\tau >0.
$$
Therefore, $g_1(y)\leq g_1(1)=-\tau,$ and so
$$
\int_0^{1}\psi(\tau(1-y))\frac{e^{\tau(\eta\log y-y)}}{y^2\sqrt{1-y}}dy \leq 
M_{\psi}e^{-\tau}\int_0^{1}\frac{1}{\sqrt{1-y}}dy =2M_{\psi}e^{-\tau}.
$$
Plugging into (\ref{I}) we obtain, as $\taurmi,$  $I(\eta, \tau) \leq C\tau^{\frac{1}{2}}e^{-\tau(\eta\log\eta-\eta+1)}$
for some positive constant $C.$ Since $\eta\log\eta-\eta+1>0$ for all $\eta>1$, we conclude that $I(\eta, \tau)\rightarrow 0$
as $\taurmi,$ and this proves the result when $\eta>1.$

Consider now the case $\eta\in (0, 1)$.

Observe first that the exponential term inside the integral in (\ref{I}) has a unique maximum at $y=\eta$.
Therefore, in order to estimate (\ref{I}) when $\taurmi$ we are going to study its behaviour
around the maximum separately from the other cases. To this end, let $\ve < \min\{\eta e^{-1},
1-\eta\}$ and write
\beqa
I(\eta, \tau) & = & \tau^{\frac{1}{2}}\eta^{-\eta\tau}e^{\tau\eta}
\left(\int_0^{\ve}+\int_{\ve}^{1-\ve}+\int_{1-\ve}^{1}\right)\psi(\tau(1-y))\frac{e^{\tau(\eta\log y-y)}}{y^2\sqrt{1-y}}dy\nn \\
& =: & I_1(\eta, \tau)+I_2(\eta, \tau)+I_3(\eta, \tau). 
\label{III}
\eeqa
The study of the integral $I_1(\eta, \tau)$ is entirely  analogous to what was done in the case $\eta>1:$ 
 since $0<y<\ve<\eta e^{-1}<\eta$ we now have, for all $\tau>\frac{2}{(1-e^{-1})\eta}$,
$$
g_1'(y) = \frac{\eta\tau-2}{y}-\tau = \frac{\tau(\eta-y)-2}{y} >0.
$$
Thus, $g_1(y)\leq g_1(\ve)<g_1(\eta e^{-1}) = (\eta\tau-2)\log\eta-(\eta\tau-2)-\tau\eta e^{-1},$ and we can write
\beqa
I_1(\eta, \tau) & \leq & \tau^{\frac{1}{2}}
e^{\tau\eta(1-\log \eta)+(\eta\tau-2)\log\eta-(\eta\tau-2)-\tau\eta/e}\Mp\int_0^{\ve}\frac{1}{\sqrt{1-y}}dy\nn \\
& \leq & C\tau^{\frac{1}{2}}e^{-\tau\eta/e} \longrightarrow 0 \quad {\rm as}\;\taurmi, \label{limI1}
\eeqa
for some positive constant $C.$

For the integral $I_3(\eta, \tau)$, define the function $g_3: (0,1 )\rightarrow \Rb$ by
$g_3(y):= (\eta\log\eta-\eta)-(\eta\log y-y).$ Since $g_3$ has a unique minimum at
$y=\eta$, where its value is $g_3(\eta)=0,$ we conclude that $y>1-\ve>\eta \Rightarrow 
g_3(y)>g_3(1-\ve)>0$ and thus, there exists a positive constant $C$ such that
\beqa
I_3(\eta, \tau) & = & \tau^{\frac{1}{2}}\int_{1-\ve}^1\psi(\tau(1-y))\frac{e^{-\tau g_3(y)}}{y^2\sqrt{1-y}}dy\nn \\
& \leq & C\tau^{\frac{1}{2}}e^{-\tau g_3(1-\ve)} \longrightarrow 0 \quad {\rm as}\;\taurmi, \label{limI3}
\eeqa
From (\ref{limI1}) and (\ref{limI3}) we conclude that 
\beq
I(\eta, \tau) = I_2(\eta, \tau) +\smallo(1) \quad {\rm as}\; \taurmi.\label{II2o}
\eeq
In order to estimate $I_2(\eta, \tau)$ we proceed as follows: we now have $\tau(1-y)>\tau\ve \rightarrow +\infty$ 
as $\taurmi,$ and thus $\psi(\tau(1-y)) = 1 +\smallo(1)$ when evaluating $I_2$ for large values of $\tau.$
As in a similar situation in Section~\ref{sec4}, we have 
$\forall\delta>0,\,\exists T(\delta):\,\forall \tau>T(\delta),\, 
\psi(\tau(1-y)) \in [1-\delta, 1+\delta]$, and we can estimate
\beq
(1-\delta)\tau^{\frac{1}{2}}\eta^{-\eta\tau}J(\eta, \tau) 
\leq I_2(\eta, \tau)\leq 
(1+\delta)\tau^{\frac{1}{2}}\eta^{-\eta\tau}J(\eta, \tau)
\quad\text{as $\taurmi,$}\label{20}
\eeq
with
$$
J(\eta, \tau) := \int_{\ve}^{1-\ve}\frac{1}{y^2\sqrt{1-y}}e^{-\tau\phi(y)}dy,
$$
and $\phi: (0, 1)\rightarrow \Rb$ is defined by $\phi(y):=y-\eta\log y -\eta.$ Since this function
is smooth and has a unique minimum, attained at $y=\eta\in (\ve, 1-\ve)$ with value 
$\phi(\eta)=-\eta\log\eta$ and $\phi''(\eta)=\eta^{-1},$ 
Laplace's method for the asymptotic evaluation of integrals 
\cite[pg 431]{af03} is applicable to $J(\eta, \tau)$, and we obtain, as $\taurmi,$
\beq
J(\eta, \tau) = e^{\tau\eta\log\eta}\frac{1}{\eta^2\sqrt{1-\eta}}\sqrt{\frac{2\pi}{\tau/\eta}}+
{\cal O}\left(\frac{e^{\tau\eta\log\eta}}{\tau^{3/2}}\right).\label{laplace}
\eeq
Hence, from (\ref{16}), (\ref{I}), (\ref{II2o}), (\ref{20}) and (\ref{laplace}), we can write, as $\taurmi,$
\beqa
\vf_1(\eta\tau, \tau) & = &\;\; \frac{1}{\sqrt{2\pi}}\eta^{\frac{3}{2}-\eta\tau}\tau^{\frac{1}{2}}
e^{\tau\eta\log\eta}\frac{1}{\eta^2\sqrt{1-\eta}}
\sqrt{\frac{2\pi}{\tau/\eta}}\left(1+{\cal O}\left(\tau^{-1}\right)\right) \label{23} \\
& &\!+\, \frac{1}{\sqrt{2\pi}}\eta^{\frac{3}{2}-\eta\tau}\tau^{\frac{1}{2}}
{\cal O}\left(\frac{e^{\tau\eta\log\eta}}{\tau^{3/2}}\right)\left(1+{\cal O}\left(\tau^{-1}\right)\right)
+\smallo(1)  \label{24} 
\eeqa
After a few trivial calculations, we imediately recognize that, as $\taurmi,$ (\ref{23}) is 
equal to $\frac{1}{\sqrt{1-\eta}}(1+\smallo(1)),$ and (\ref{24}) is 
$\tau^{-1}{\cal O}(1)\left(1+{\cal O}\left(\tau^{-1}\right)\right)+\smallo(1) ={\cal O}\left(\tau^{-1}\right)
+\smallo(1)=\smallo(1).$ 
Therefore, for $\eta\in (0, 1),$
$$
\vf_1(\eta\tau, \tau)\longrightarrow \frac{1}{\sqrt{1-\eta}}\quad\text{as $\taurmi$}
$$
and this concludes the proof in the case of monomeric initial data.

\subsection{Non monomeric initial data} 

If the initial data has
non zero components $c_k(0)$ with $k\geq 2,$ the proof of the stated similarity behaviour of 
$\tilde{c}_j(\tau)$ requires, according to (\ref{12}), that we now prove that
$$
\lim_{\substack{j,\,\tau \rightarrow +\infty\\ \eta=j/\tau\; {\rm fixed}\\ \eta\neq 1}}
\sqrt{\frac{2}{\alpha}\tau\,}\;e^{-\tau}\sum_{k=2}^{j}\frac{\tau^{j-k}}{(j-k)!}c_k(0)
=0. 
$$
Define $\nu := \eta^{-1}$, write $\tau = j\nu,$ and use the assumption on the initial 
condition, namely $c_k(0)\leq\rho/k^{\mu}.$ We thus have
\beqa
\sqrt{\frac{2}{\alpha}\tau\,}\;e^{-\tau}\sum_{k=2}^{j}\frac{\tau^{j-k}}{(j-k)!}c_k(0) & \leq &
\rho\sqrt{\frac{2}{\alpha}\,}\sqrt{j\nu}\;e^{-j\nu}\sum_{k=2}^{j}\frac{(j\nu)^{j-k}}{(j-k)!k^{\mu}}\nn \\
& =: &  \rho\sqrt{\frac{2}{\alpha}\,}\vf_2(\nu, j)\nn
\eeqa
where $\vf_2$ is defined by the equality. Our goal is to prove that $\vf_2(\nu, j)\rightarrow
0$ as $\jri$, for all positive $\nu \neq 1.$ 

Like it was done before, we shall study the cases $\nu>1$ and $\nu<1$ separately. By the
heuristic geometric reasons explained in the Introduction, the case $\nu>1$ ($\eta<1$) is now easier
to handle. In fact, for this case, even using the notoriously bad upper
bound to the sum provided by taking the product of the number of terms by their maximum, we
only need to impose $\mu\geq 0$. Using this same approach to estimate $\vf_2$ in the
case $\nu<1$ ($\eta>1$) we would need to impose $\mu>1,$ which would not even guarantee that
all finite mass initial data are included. So, for this second part of the proof, a finer approach
is required, which consists in estimating the small and the large $\ell$ contributions to
$\vf_2$ separately.

Consider first the case $\nu>1.$

Change the summation variable $k\mapsto \ell:=j-k.$ It is sufficient, for this range of
$\nu$, to  bound $\vf_2$ as follows
\beq
\vf_2(\nu, j) = \sqrt{j\nu}\;e^{-j\nu}\sum_{\ell=0}^{j-2}\frac{(j\nu)^{\ell}}{\ell!(j-\ell)^{\mu}}\leq
\frac{1}{2^{\mu}} \sqrt{j\nu}\;e^{-j\nu}\sum_{\ell=0}^{j-2}\frac{(j\nu)^{\ell}}{\ell!}. \label{*}
\eeq
Considering the sequence $u_{\ell}:= \frac{(j\nu)^{\ell}}{\ell!},$ and
studying the sign of
$$u_{\ell+1}-u_{\ell}=\frac{(j\nu)^{\ell}}{\ell!}\left(\frac{j\nu}{\ell+1}-1\right),$$
we conclude the maximum of $u_{\ell}$ is attained at $\ell = \lfloor j\nu\rfloor >
j\nu-1>j-1>j-2.$ As $\jri$, we can thus estimate the right-hand side of (\ref{*}):
\beqa
\frac{1}{2^{\mu}} \sqrt{j\nu}\;e^{-j\nu}\sum_{\ell=0}^{j-2}\frac{(j\nu)^{\ell}}{\ell!} 
& < & \frac{1}{2^{\mu}} \sqrt{j\nu}\;e^{-j\nu}(j-1)\frac{(j\nu)^{j-2}}{(j-2)!} \nn \\
& = & \frac{1}{2^{\mu}} (j\nu)^{-3/2}\;e^{-j\nu}(j-1)^2\frac{(j\nu)^{j}}{(j-1)!}\nn \\
& = & \frac{\nu^{-3/2}}{2^{\mu}\sqrt{2\pi}} j^{-3/2}\;e^{-j\nu}(j-1)^2
      \frac{(j\nu)^{j}}{e^{-j}j^jj^{-1/2}}(1+\smallo(1))\nn \\
& = & \frac{\nu^{-3/2}}{2^{\mu}\sqrt{2\pi}}\left(\frac{j-1}{j}\right)^2e^{-j(\nu-1-\log\nu)+\log j}(1+\smallo(1)) \nn \\
&   & \label{**}
\eeqa  
where we use Stirling's approximation in the second equality. Since $\nu >1,$ we have $\nu -1-\log\nu>0$, 
and so $-j(\nu-1-\log\nu)+\log j \rightarrow -\infty$ as $\jri$ from which we conclude (\ref{**})
goes to zero as $\jri,$ and we obtain the result we seek in the case $\nu >1.$

Consider now the case $\nu \in (0, 1).$

Let $\beta\in \left(\nu e^{1-\nu}, \min\{\nu e, 1\}\right)$ be fixed, and write
\beq
\vf_2(\nu, j) = \sqrt{j\nu}\;e^{-j\nu}\left(\sum_{0\leq\ell\leq\beta j}\frac{(j\nu)^{\ell}}{\ell!(j-\ell)^{\mu}}
+\sum_{\beta j<\ell\leq j-2}\frac{(j\nu)^{\ell}}{\ell!(j-\ell)^{\mu}}\right) =: S_1(j)+S_2(j)    \nn
\eeq
For the first sum we have
\beqa
S_1(j) & \leq & \sqrt{j\nu}\;e^{-j\nu}\sum_{0\leq\ell\leq\beta j}\frac{(j\nu)^{\ell}}{\ell!(j-\beta j)^{\mu}} \nn \\
& = & \sqrt{j\nu}\;e^{-j\nu}\frac{1}{j^{\mu}(1-\beta)^{\mu}}\sum_{0\leq\ell\leq\beta j}\frac{(j\nu)^{\ell}}{\ell!}\nn \\
& \leq & \sqrt{j\nu}\;e^{-j\nu}\frac{1}{j^{\mu}(1-\beta)^{\mu}}e^{j\nu} \nn \\
& = & 
\frac{\sqrt{\nu}}{(1-\beta)^{\mu}}j^{\frac{1}{2}-\mu} \longrightarrow 0 \quad\mbox{as}\;\,\jri,\;\,\mbox{if}\;\,\mu>\frac{1}{2}.\nn
\eeqa
By Stirling's expansion we can write, for all sufficiently large $\ell,$ $\ell!\geq e^{-\ell}\ell^{\ell+\frac{1}{2}}.$ Therefore,
since in $S_2$ we have $(j-\ell)^{\mu}\geq 2^{\mu}\geq 1,$ we can estimate that term as follows,
\beqa
S_2(j) & \leq & \sqrt{j\nu}\;e^{-j\nu}\sum_{\beta j<\ell\leq j-2}\frac{(j\nu)^{\ell}}{e^{-\ell}\ell^{\ell+\frac{1}{2}}}\nn \\
& = & \sqrt{j\nu}\;e^{-j\nu}\sum_{\beta j<\ell\leq j-2}\ell^{-\frac{1}{2}}\left(\frac{j\nu e}{\ell}\right)^{\ell}\nn \\
& < & \sqrt{\frac{\nu}{\beta}}\;e^{-j\nu}\sum_{\beta j<\ell\leq j-2}\left(\frac{j\nu e}{\ell}\right)^{\ell}\nn \\
& < & \sqrt{\frac{\nu}{\beta}}\;e^{-j\nu}\frac{1-\left(\frac{\nu e}{\beta}\right)^{j-2-\lfloor\beta j\rfloor}}
{1-\frac{\nu e}{\beta}}\left(\frac{\nu e}{\beta}\right)^{\lfloor\beta j\rfloor+1} \nn \\
& < & \frac{\sqrt{\nu\beta}}{\nu e-\beta}\;e^{-j\nu}\left(\frac{\nu e}{\beta}\right)^{j-1} \nn \\
& = & \frac{\sqrt{\nu\beta}}{\nu e-\beta}\frac{\beta}{\nu e}\left(\frac{\nu e^{1-\nu}}{\beta}\right)^{j}  
\longrightarrow 0 \quad\mbox{as}\;\,\jri.\nn 
\eeqa

This completes the proof of the theorem.  \hfill\qed

\section{Self-similar behaviour of the coagulation system along the characteristic direction}   
\label{sec6}

Consider the function $\Phi_2:\Rb\rightarrow\Rb$ defined by 
$$
\Phi_2(\xi):= e^{-\frac{1}{2}\xi^2}\!\int_0^{+\infty}\!\!e^{-\xi w^2-\frac{1}{2}w^4}dw.
$$
The main result of this part of the paper, complementing Theorem~\ref{mainteo1}, is the following

\subsection{Monomeric initial data} 


\begin{theo} \label{mainteo2} 
Let $(c_j)$ be any non-negative solution of (\ref{1}) with monomeric initial data.
Let $\tau(t)$ and $\tilde{c}_j(\tau)$ be as given in (\ref{eq:tau}) and (\ref{eq:ctilde}) respectively.
Then,
$$
\lim_{\substack{j,\,\tau \rightarrow +\infty\\ \xi=\frac{j-\tau}{\sqrt{\tau}}\; {\rm fixed}\\
\xi\in\Rb}}
\left(\frac{\pi^2}{\alpha^2}\tau\!\right)^{1/4}\!
\tilde{c}_j(\tau) = \Phi_2(\xi).
$$
\end{theo}




\vspace*{4mm}
\noindent
After a few manipulations with the integral defining $\Phi_2$ we can write
$$
\Phi_2(\xi)= \frac{\pi}{4}\sqrt{|\xi|}e^{-\frac{1}{4}\xi^2}\left(I_{-\frac{1}{4}}\left(\tfrac{1}{4}\xi^2\right)
+\sgn(-\xi)I_{\frac{1}{4}}\left(\tfrac{1}{4}\xi^2\right)\right),
$$
where $I_{\nu}(\cdot)$ is the modified Bessel function \cite[pp 374-7]{AS}, and the value at
$\xi=0$ is defined to be the limit of the right-hand side as $\xi\rightarrow 0.$
Actually, $\Phi_2$ can be seen as an element of a larger family of functions: it is the
function obtained by making $\nu=\frac{1}{2}$ in
$$
\begin{array}{l}
g_{\nu}(\xi) :=\\
:= \frac{\Gamma(1-\tfrac{1}{2}\nu)|\xi|e^{-\frac{1}{2}\xi^2}}{2^{\nu/2-1/2}\sqrt{\pi}}\left(
\frac{\Gamma\left(\tfrac{1}{2}-\tfrac{1}{2}\nu\right)}{4\sqrt{\pi}}
U\left(1-\tfrac{1}{2}\nu, \tfrac{3}{2}; \tfrac{1}{2}\xi^2\right)
+\tfrac{\sgn(-\xi)+1}{2}M\left(1-\tfrac{1}{2}\nu, \tfrac{3}{2}; \tfrac{1}{2}\xi^2\right)\right).
\end{array}
$$
These functions were formally deduced in \cite{w04} as similarity profiles for 
addition models with time dependent monomer inputs $J_0(t)= \alpha t^{\omega},$
where $\nu= -\frac{\omega-1}{\omega+2}.$ The functions $U$ and $M$ are 
Kummer's hypergeometric functions \cite[pp 504-515]{AS}.

\vspace*{4mm}
\noindent
{\bf{Proof}}\\

\noindent
For monomeric initial data (\ref{12}) reduces to
$$
\left(\frac{\pi^2}{\alpha^2}\tau\!\right)^{1/4}\!\tilde{c}_j(\tau) =
\frac{\left(\frac{\pi^2}{\alpha^2}\tau\!\right)^{1/4}}{(j-2)!}
\int_0^{\tau}\tilde{c}_1(\tau-s)s^{j-2}e^{-s}ds.
$$
Consider the function $\vf_3$ defined in $[2, \infty)\times[0, \infty)$ by
\beq
\vf_3(x, \tau) := \frac{\left(\frac{\pi^2}{\alpha^2}\tau\!\right)^{1/4}}{\Gamma(x-1)}
\int_0^{\tau}\tilde{c}_1(\tau-s)s^{x-2}e^{-s}ds. \label{fi}
\eeq
In order to consider the similarity limit we introduce the variable
$\xi := \frac{j-\tau}{\sqrt{\tau}}$ and write (\ref{fi}) in the similarity
variable $\xi$ as follows:
\beq
\simf := \frac{\left(\frac{\pi^2}{\alpha^2}\tau\!\right)^{1/4}}
{\Gamma(\tau+\xi\sqrt{\tau}-1)}
\int_0^{\tau}\tilde{c}_1(\tau-s)s^{\tau+\xi\sqrt{\tau}-2}e^{-s}ds. \label{simfi}
\eeq
Since $\vf_3(j,\tau) = \left(\frac{\pi^2}{\alpha^2}\tau\!\right)^{1/4}\!\tilde{c}_j(\tau)$
if $2\leq x=j\in \Nb$, to prove the theorem it suffices to show that
\beq
\lim_{\tau\rightarrow+\infty}\simf = 
\Phi_2(\xi). \label{limsimfi}
\eeq
We start by considering, in the integral in the right-hand side of (\ref{simfi}), the change of
variables
$$s \mapsto w:= \sqrt{\sqrt{\tau}-\frac{s}{\sqrt{\tau}}}$$
and by writing $\psi(\cdot) := \sqrt{\frac{2}{\alpha}(\cdot)}\;\tilde{c}_1(\cdot).$ This
results in
$$
\simf = \frac{2\pi}{\Gamma(\tau+\xi\sqrt{\tau}-1)}\int_0^{\tau^{1/4}}
\!\!\!\!\!\!\!\!\!\psi(\sqrt{\tau}w^2)\sqrt{\tau}
(\tau-\sqrt{\tau}w^2)^{\tau+\xi\sqrt{\tau}-2}e^{-\tau+\sqrt{\tau}w^2}dw.
$$
From Stirling's expansion for the Gamma function we have, as $\taurmi$,
\beq
\simf = \frac{\tau^{\tau+\xi\sqrt{\tau}-\frac{3}{2}}(1+\smallo(1))}
{e^{-\xi\sqrt{\tau}}(\tau+\xi\sqrt{\tau})^{\tau+\xi\sqrt{\tau}-\frac{3}{2}}}
\int_0^{\tau^{1/4}}\!\!\!\!\!\!\!\!\!\psi(\sqrt{\tau}w^2)
\left(1-\frac{w^2}{\sqrt{\tau}}\right)^{\tau+\xi\sqrt{\tau}-2}\!\!\!\!\!\!\!
e^{\sqrt{\tau}w^2}dw
\label{simfistir}
\eeq
We can now estimate the behaviour as $\taurmi$ of the  multiplicative factor in the 
right-hand side of (\ref{simfistir}) as follows:
\beqa
\lefteqn{\frac{\tau^{\tau+\xi\sqrt{\tau}-\frac{3}{2}}}
{e^{-\xi\sqrt{\tau}}(\tau+\xi\sqrt{\tau})^{\tau+\xi\sqrt{\tau}-\frac{3}{2}}}=}\nn \\
& = & e^{\xi\sqrt{\tau}}\left(
\frac{\tau}{\tau+\xi\sqrt{\tau}}\right)^{\tau+\xi\sqrt{\tau}}
\left(1+\frac{\xi}{\sqrt{\tau}}\right)^{3/2}\nn \\
& = & e^{\xi\sqrt{\tau}}\frac{1}{\left(1+\frac{\xi}{\sqrt{\tau}}\right)^{\tau}}
\frac{1}{\left(\left(1+\frac{\xi}{\sqrt{\tau}}\right)^{\sqrt{\tau}}\right)^{\xi}}(1+\smallo(1)) \nn \\
&=& \left(\frac{e}{\left(1+\frac{\xi}{\sqrt{\tau}}\right)^{\sqrt{\tau}/\xi}}
\right)^{\xi\sqrt{\tau}}\!\!\!\!\!\!e^{-\xi^2}(1+\smallo(1))\nn \\
& = & e^{\frac{1}{2}\xi^2}e^{-\xi^2}(1+\smallo(1)) \, = \, e^{-\frac{1}{2}\xi^2}(1+\smallo(1))\nn
\eeqa
where the last but one equality is obtained by changing variable 
$\tau\mapsto x:=\frac{\xi}{\sqrt{\tau}}$
and applying L'H\^opital's rule twice. Thus, we can write (\ref{simfistir}), in the $\taurmi$ limit, as
\beq
\simf = e^{-\frac{1}{2}\xi^2}(1+\smallo(1))
\int_0^{\tau^{1/4}}\!\!\!\!\!\!\!\!\!\psi(\sqrt{\tau}w^2)
e^{(\tau+\xi\sqrt{\tau}-2)\log\left(1-\frac{w^2}{\sqrt{\tau}}\right)+\sqrt{\tau}w^2}
\!\!\!dw.
\label{simfiexp}
\eeq
In order to estimate the integral, and considering that the only relevant information available
about $\psi(s)$ is that it is a non-negative, continuous, and bounded function on $[0, \infty)$
and $\psi(s)=1+\smallo(1)$ as $s\rightarrow +\infty$, we are forced to treat the region with
$w$ close to zero separately. The idea is to write the integral as $\int_0^{\ve} + 
\int_{\ve}^{\tau^{1/4}}$ and to prove the first integral can be made arbitrarily small. More 
precisely, we now prove the following:
\beq
\forall_{\delta>0},
\exists_{\substack{\ve_0=\ve_0(\xi)\\ \tau_0=\tau_0(\xi, \ve_0)}}:
\forall_{\substack{\ve<\ve_0\\ \tau>\tau_0}},\;
\int_0^{\ve}\!\!\psi(\sqrt{\tau}w^2)
e^{(\tau+\xi\sqrt{\tau}-2)\log\left(1-\frac{w^2}{\sqrt{\tau}}\right)+\sqrt{\tau}w^2}
\!\!\!dw < \delta.\label{smallve}
\eeq
In fact, let $\delta>0$ be arbitrary, let $\ve_0=\ve_0(\xi) := \max\left\{\ve: 
  e^{|\xi|\ve^2}- \frac{\delta}{M_{\psi}e}\frac{1}{\ve}\leq 0\right\},$ with
$M_{\psi} = \displaystyle{\sup_{s\in[0, \infty)}\psi(s)}$, and let 
$\tau_0:=\max\left\{4\ve_0^4, \frac{1}{4}(\sqrt{\xi^2+8}-\xi)^{2}\right\}$. 
Then, for all $\tau>\tau_0$  we have $\tau+\xi\sqrt{\tau}-2>0,$ and,using 
$\log(1-x)\leq -x$,  we conclude that, for $\ve<\ve_0$,
\beqa
\lefteqn{\int_0^{\ve}\!\!\psi(\sqrt{\tau}w^2)
e^{(\tau+\xi\sqrt{\tau}-2)\log\left(1-\frac{w^2}{\sqrt{\tau}}\right)+\sqrt{\tau}w^2}
dw \leq }\nn \\
& \leq & M_{\psi}\int_0^{\ve}\!\!
e^{(\tau+\xi\sqrt{\tau}-2)\log\left(1-\frac{w^2}{\sqrt{\tau}}\right)+\sqrt{\tau}w^2}
dw  \nn \\
& \leq & M_{\psi}\int_0^{\ve}\!\!e^{-\xi w^2}e^{\frac{2w^2}{\sqrt{\tau}}}dw  \nn \\
& \leq & M_{\psi}\ve e^{|\xi|\ve^2}e^{\frac{2\ve^2}{\sqrt{\tau}}}  \nn \\
& < & M_{\psi}e\frac{\delta}{M_{\psi}e} \, = \, \delta.\nn
\eeqa
We are now left with the contribution $\int_{\ve}^{\tau^{1/4}}$ to the integral in (\ref{simfiexp}).
Since $w\geq\ve\Rightarrow\sqrt{\tau}w^2 \geq \sqrt{\tau}\ve^2 \rightarrow +\infty$ as $\taurmi,$
it is easy to see that, for all sufficiently large $\tau$ we have the integral of interest
asymptotically equal to 
\beq
(1+\smallo(1))\int_{\ve}^{\tau^{1/4}}\!\!
e^{(\tau+\xi\sqrt{\tau}-2)\log\left(1-\frac{w^2}{\sqrt{\tau}}\right)+\sqrt{\tau}w^2}
dw.\label{intlargetau}
\eeq
To estimate (\ref{intlargetau}) observe that, as $\taurmi,$
\beqa
\lefteqn{e^{(\tau+\xi\sqrt{\tau}-2)\log\left(1-\frac{w^2}{\sqrt{\tau}}\right)+\sqrt{\tau}w^2} = }\nn\\
& = & e^{\sqrt{\tau}w^2}\left(1-\frac{w^2}{\sqrt{\tau}}\right)^{\tau}
\left(\left(1-\frac{w^2}{\sqrt{\tau}}\right)^{\sqrt{\tau}}\right)^{\xi}
\left(1-\frac{w^2}{\sqrt{\tau}}\right)^{-2}   \nn \\
& = & \left(\frac{\left(1-\frac{w^2}{\sqrt{\tau}}\right)^{\sqrt{\tau}}}{e^{-w^2}}\right)^{\sqrt{\tau}}
e^{-\xi w^2}(1+\smallo(1))\nn\\
& = & e^{-\xi w^2-\frac{1}{2}w^4}(1+\smallo(1))
\eeqa
where the last equality is obtained by changing the variable $\tau\mapsto x:=\frac{1}{\sqrt{\tau}}$
and applying L'H\^opital's rule. From this we conclude that there exists a continuous function
$g(w; \tau)$ defined for $\tau>\ve^4,$ $w\in[\ve, \tau^{1/4})$, satisfying $1+g(w; \tau)\geq 0$ and 
$g(w; \tau) \rightarrow 0$ as $\taurmi$ for each fixed $w,$ such that
\beq
e^{(\tau+\xi\sqrt{\tau}-2)\log\left(1-\frac{w^2}{\sqrt{\tau}}\right)+\sqrt{\tau}w^2} = 
e^{-\xi w^2-\frac{1}{2}w^4}(1+g(w; \tau)).\label{gdef}
\eeq
Considering $\tau$ sufficiently large, for instance larger than $\tau_0$ previously defined, we
can use the bound on the logarithm to obtain, from (\ref{gdef}),
\beqa
\lefteqn{1+g(w; \tau) = }\nn \\
& = &\exp\left((\tau+\xi\sqrt{\tau}-2)\log\left(1-\frac{w^2}{\sqrt{\tau}}\right)+
\sqrt{\tau}w^2+\xi w^2+\frac{1}{2}w^4\right) \nn \\
& \leq & \exp\left((\tau+\xi\sqrt{\tau}-2)\left(-\frac{w^2}{\sqrt{\tau}}-\frac{w^4}{2\tau}
-\frac{w^6}{3\tau^{3/2}}\right)+
\sqrt{\tau}w^2+\xi w^2+\frac{1}{2}w^4\right) \nn \\
& = &  \exp\left(2\frac{w^2}{\sqrt{\tau}}-\frac{\xi}{2}\frac{w^4}{\sqrt{\tau}}+
\frac{w^4}{\tau}-\frac{1}{3}(\tau+\xi\sqrt{\tau}-2)\frac{w^6}{\tau^{3/2}}\right)\nn \\
& =: & e^{h(w; \tau)}\nn
\eeqa
where the last equality defines $h.$ Change variables $(w, \tau) \mapsto (v, \theta)$ where
$v:=\frac{w^2}{\sqrt{\tau}}$ and $\theta=\sqrt{\tau}.$ Denote by $\tilde{h}$ the function $h$ in 
the new variables. We immediately conclude that $\tilde{h}$ is the polynomial
$$
\tilde{h}(v; \theta) = 2v -\frac{\xi}{2}v^2\theta +v^2-\frac{1}{3}(\theta^2+\xi\theta-2)v^3
$$
and the region of interest in the new variables is $A:=\{(v, \theta): \theta>\ve^2,\,  
\frac{\ve^2}{\theta}\leq v <1\}.$ Clearly $\tilde{h}$ is bounded from above in $A$. Denote by 
$\log M$ an upper bound for $\tilde{h}$. Thus, for all $\tau>\tau_0$ we can write
$1+g(w; \tau)\leq M.$
Since $g(w; \tau) \rightarrow 0$ as $\taurmi$ for each fixed $w,$ the quantity 
$$\tilde{\tau}_w := \max\left\{\tau: |g(w; \tau)|\geq \delta \right\}$$ 
is well defined
for each fixed $w$ and, because $g$ is continuous, 
$$\tau_1:= \max\left\{\tau_0, \textstyle{\sup_{w\in[\ve, w_0]}}\tilde{\tau}_w\right\}<+\infty.$$
Two further auxiliary estimates, both immediate, are needed: firstly, knowing that 
$e^{-\xi w^2-\frac{1}{2}w^4}\in L^1(\Rb)$, we can define  
$$
{\cal I}(\xi, w_0):= \int_{w_0}^{+\infty}e^{-\xi w^2-\frac{1}{2}w^4}dw,
$$
and conclude that
\beq
\forall_{\delta>0} \exists_{\tilde{w}_0(\xi)}: \forall_{w_0>\tilde{w}_0}, \;\;
{\cal I}(\xi, w_0) < \delta,\label{resto}
\eeq
and secondly, given $\delta>0$ arbitrarily, and defining $\ve_1= \ve_1(\xi)$ by the
expression $\ve_1 :=\max\left\{\ve: e^{|\xi|\ve^2}- \frac{\delta}{\ve}\leq 0\right\},$ 
it is obvious that,
\beq
\forall_{\delta>0} \exists_{\ve_1}: \forall_{\ve<\ve_1}, \;\;
\int_0^{\ve}e^{-\xi w^2}dw < \delta. \label{origem}
\eeq
From the above estimates, and using the notation already introduced, we deduce that:
$$
\forall_{\delta>0},
\exists_{\substack{\tilde{\ve}=\min\{\ve_0, \ve_1\}\\ \tau_0=\tau_0(\xi, \ve_0)}}: 
\forall_{\ve<\tilde{\ve}}, 
\exists_{\tilde{w}=\max\{\ve, \tilde{w}_0\}}: 
\forall_{w_0>\tilde{w}}, 
\exists_{\tau_1=\tau_1(\tau_0, w_0, \ve)}:  
\forall_{\tau>\tau_1},
$$
\beqa
\lefteqn{\int_{\ve}^{\tau^{1/4}}(1+g(w; \tau))e^{-\xi w^2-\frac{1}{2}w^4}dw  =} \nn \\
& = & \int_{\ve}^{w_0}(1+g(w; \tau))e^{-\xi w^2-\frac{1}{2}w^4}dw +
\int_{w_0}^{\tau^{1/4}}(1+g(w; \tau))e^{-\xi w^2-\frac{1}{2}w^4}dw \nn \\
& \leq & (1+\delta)\int_{\ve}^{w_0}e^{-\xi w^2-\frac{1}{2}w^4}dw +M {\cal I}(\xi, w_0) \nn \\
& \leq & \int_{0}^{+\infty}e^{-\xi w^2-\frac{1}{2}w^4}dw +({\cal I}(\xi, 0)+M)\delta \nn \\
& \leq & \int_{0}^{+\infty}e^{-\xi w^2-\frac{1}{2}w^4}dw +C\delta \nn
\eeqa
and also, using the positivity of $1+g(w; \tau),$
\beqa
\lefteqn{\int_{\ve}^{\tau^{1/4}}(1+g(w; \tau))e^{-\xi w^2-\frac{1}{2}w^4}dw  \geq }\nn \\
& \geq & \int_{\ve}^{w_0}(1+g(w; \tau))e^{-\xi w^2-\frac{1}{2}w^4}dw \nn \\
& \geq & (1-\delta)\int_{\ve}^{w_0}e^{-\xi w^2-\frac{1}{2}w^4}dw \nn \\
& = & \int_{0}^{+\infty}e^{-\xi w^2-\frac{1}{2}w^4}dw - 
\left(\int_0^{\ve}+\int_{w_0}^{\infty}+\delta\int_{\ve}^{w_0}\right)e^{-\xi w^2-\frac{1}{2}w^4}dw\nn \\
& \geq & \int_{0}^{+\infty}e^{-\xi w^2-\frac{1}{2}w^4}dw - \left({\cal I}(\xi, 0)+2\right)\delta \nn\\
& \geq & \int_{0}^{+\infty}e^{-\xi w^2-\frac{1}{2}w^4}dw - C\delta \nn
\eeqa
where $C:= {\cal I}(\xi, 0)+\max\{2, M\}.$

Plugging this result into (\ref{intlargetau}) and remembering (\ref{smallve}) we 
finally conclude (\ref{limsimfi}). This concludes the proof.  \hfill\qed  

\subsection{Remarks concerning non-monomeric initial data} 

It is our conviction that Theorem~\ref{mainteo2} also holds for non-monomeric initial data
satisfying the decay condition of Theorem~\ref{mainteo1}. A few numerical runs gave results consistent
with this conviction. However, we were not yet able to prove this is indeed so.
In the present final section we shall describe what was achieved so far, and what were the difficulties 
encountered.
 
From expression (\ref{12}), if we consider non-monomeric initial data in Theorem~\ref{mainteo2},
we need to study the value of the limit
\beq
\lim_{\substack{j,\,\tau \rightarrow +\infty\\ \xi=\frac{j-\tau}{\sqrt{\tau}}\; {\rm fixed}}}
\left(\frac{\pi^2}{\alpha^2}\tau\!\right)^{1/4}\!e^{-\tau}\sum_{k=2}^{j}\frac{\tau^{j-k}}{(j-k)!}c_k(0).\label{criticallimit}
\eeq
Using the definition of $\xi$ we can write $j=\tau+\xi\sqrt{\tau},$ which, solved for $\tau$ gives
$\tau = j\Delta_j$ where the function $\Delta_j:\Rb\longrightarrow\Rb$ is defined by
\beq
\Delta_j(\xi):=\left(\sqrt{1+\frac{\xi^2}{4j}}\;+\sgn(-\xi)\sqrt{\frac{\xi^2}{4j}}\;\right)^2. \label{deltaj}
\eeq
A number of properties of this function can be easily established, in particular,
\begin{align}
\Delta_j & \,\left\{
\begin{aligned}
<1 & \text{\;\;if $\xi>0$}\\ =1 & \text{\;\;if $\xi=0$} \\ >1 & \text{\;\;if $\xi<0$,}
\end{aligned}\right. \label{signdelta}\\
\Delta_j  & \rightarrow 1 \text{\;\;as $\jri$, for all $\xi$,} \label{limitdelta}\\
\intertext{and the following power series expansion}
\Delta_j & = 1-\frac{1}{\sqrt{j}}\xi + \frac{1}{2j}\xi^2 +\sum_{n=1}^{\infty}(-1)^{n-1}
\frac{(2n-3)!!}{n!2^{2n+2}}\left(\frac{\xi}{\sqrt{j}}\right)^{2n+1}, \text{\; for $j>\frac{1}{4}\xi^2.$}\label{deltaseries}
\end{align}
Assuming $c_k(0)\leq \rho/k^{\mu}$, changing the summation variable to $\ell:=j-k$, writing
$\tau$ in terms of $j$ as indicated, and not taking multiplication constants
into account, the limit (\ref{criticallimit}) reduces to
\beq
\lim_{\jri}
j^{1/4}\!e^{-j\Delta_j}\sum_{\ell=0}^{j-2}\frac{(j\Delta_j)^{\ell}}{\ell !(j-\ell)^{\mu}}.\label{jlimit}
\eeq
which corresponds to the study of $\varphi_2$ in the proof of Theorem~\ref{mainteo1}. The fact that now
$\Delta_j$ is not bounded away from $1$, as was the case with $\nu$ in the above mentioned proof, is
the origin of the main difficulties. The most promising approach to the
evaluation of (\ref{jlimit}) is to proceed by decomposing the sum into a small $\ell$
and a large $\ell$ contribution, similar to what was done in the case $\nu\in (0, 1).$ In that occasion the 
cut off size separating the two sums was at $\ell =\beta j$ for a $\beta$ fixed in a given way.
This cut off scales like $\tau$ as a function of $j.$ It is natural to keep considering a 
cut off fulfilling the same scaling requirement, which in this case means $\sim j-|\xi|\sqrt{j}.$
So let us define $j^{\star}:=(j-2)-(1+|\xi|)\sqrt{j-2}$ and write the expression in (\ref{jlimit})
as
\beq
j^{1/4}\!e^{-j\Delta_j}\left(\sum_{0\leq\ell\leq j^{\star}}\frac{(j\Delta_j)^{\ell}}{\ell !(j-\ell)^{\mu}}
+\sum_{j^{\star}<\ell\leq j-2}\frac{(j\Delta_j)^{\ell}}{\ell !(j-\ell)^{\mu}}\right) =: S_3(j)+S_4(j)    \nn
\eeq

The small $\ell$ sum, $S_3$, that corresponds to the contribution of {\it large\/} cluster
in the initial data (remember the change of variable $k\mapsto\ell$), can be estimated in the same way as
was done for the sum $S_1$ in the proof of Theorem~\ref{mainteo1}:
\beqa
S_3(j) & \leq & j^{1/4}\;e^{-j\Delta_j}\sum_{0\leq\ell\leq j^{\star}}\frac{(j\Delta_j)^{\ell}}{\ell!(j-j^{\star})^{\mu}} \nn \\
& = & \frac{j^{\frac{1}{4}-\frac{\mu}{2}}}{\left(\frac{2}{\sqrt{j}}+(1+|\xi|)\left(1-\frac{2}{j}\right)^{1/2}\right)^{\mu}}
\;e^{-j\Delta_j}\sum_{0\leq\ell\leq j^{\star}}\frac{(j\Delta_j)^{\ell}}{\ell!}\nn \\
& \leq & C j^{\frac{1}{4}-\frac{\mu}{2}}e^{-j\Delta_j}\sum_{\ell=0}^{\infty}\frac{(j\Delta_j)^{\ell}}{\ell!} \;=\;
C j^{\frac{1}{4}-\frac{\mu}{2}}  \longrightarrow 0 \quad\mbox{as}\;\,\jri,\;\,\mbox{if}\;\,\mu>\frac{1}{2}.\nn
\eeqa

The large $\ell$ sum, $S_4$, corresponding to the contribution of {\it small\/} clusters, is the one that could not yet be tackled rigorously and
we must at present leave it as a 

\begin{conj}
With the definitions above it holds that $S_3(j)+S_4(j)\rightarrow 0$ as $\jri$, when $\mu>\frac{1}{2}$,
and hence Theorem~\ref{mainteo2} also holds for non-monomeric initial data with this decay.
\end{conj}

\vspace*{4mm}



\begin{thebibliography}{}

\bibitem{AS} M. Abramowitz, I.A. Stegun, Handbook of Mathematical Functions, Dover, New York, 1972

\bibitem{af03} M.J. Ablowitz, A.S. Fokas, Complex Variables, 2nd Ed,
Cambridge Texts in Applied Mathematics, Cambridge University Press, Cambridge 2003.

\bibitem{b94} G.S. Bales, D.C. Chrzan, Dynamics of irreversible
island growth during submonolayer epitaxy, 
Phys. Rev. B, \textbf{50}, (1994) 6057--6067. 
 
\bibitem{bcp} J.M. Ball, J. Carr, O. Penrose,  The \BD\ cluster 
equations: basic properties and asymptotic behaviour of solutions, 
Comm. Math. Phys., \textbf{104}, (1986) 657--692. 

\bibitem{be96} M.C. Bartelt, J.W. Evans, Exact island-size distributions for
submonolayer deposition: influence of correlations between island size
and separation, Phys. Rev. B, \textbf{54}, (1996) R17359--R17362. 

\bibitem{bd} R. Becker, W. D\"oring, Kinetische Behandlung in \"ubers\"attigten D\"ampfern,  
Ann. Phys. (Leipzig), \textbf{24}, (1935) 719--752. 

\bibitem{BK} N.V. Brilliantov, P.L. Krapivsky, Non-scaling and source-induced scaling behaviour in 
aggregation models of movable monomers and immovable clusters, 
J. Phys. A; Math. Gen., \textbf{24}, (1991) 4787--4803. 

\bibitem{c81} J. Carr, Applications of Centre Manifold Theory,
Applied Mathematical Sciences vol. 35, Springer-Verlag, New York, 1981 

\bibitem{dc98} F.P. da Costa, A finite-dimensional dynamical model for 
gelation in coagulation processes,
J. Nonlinear Sci., \textbf{8}, (1998) 619--653.

\bibitem{dC96} F.P. da Costa, On the dynamic scaling behaviour of solutions 
to the discrete Smoluchowski equation,
Proc. Edinburgh Math. Soc., \textbf{39}, (1996) 547--559.

\bibitem{dkw} S.C. Davies, J.R. King and J.A.D. Wattis,  
The Smoluchowski coagulation equations with continuous injection,
J. Phys. A; Math. Gen., \textbf{32}, (1999) 7745--7763. 

\bibitem{emrr} M. Escobedo, S. Mischler, M. Rodriguez-Ricard,
On self-similarity and stationary problems for fragmentation and 
coagulation models, Ann. Inst. H. Poincar\'e Anal. Non Lin\'eaire, \textbf{22}, 
(2005) 99--125. 

\bibitem{e96} J.W. Evans, M.C. Bartelt,
Nucleation, growth and kinetic roughening of metal (100) homoepitaxial
thin films, Langmuir, \textbf{12}, (1996) 217--229. 

\bibitem{FL04} N. Fournier, Ph. Lauren\c{c}ot, Existence of self-similar 
solutions to Smoluchowski's coagulation equation,
Commun. Math. Phys., \textbf{256}, (2005) 589--609.

\bibitem{grc} F. Gibou, C. Ratsch, R. Caflisch,
Capture numbers in rate equations and scaling laws for epitaxial growth,  
Phys. Rev. B, \textbf{67}, (2003) 155403. 

\bibitem{HE} E.M. Hendriks, M.H. Ernst, Exactly soluble addition 
and condensation models in coagulation kinetics, 
J. Coll. Int. Sci., \textbf{97}, (1984) 176--194.

\bibitem{kw} J.R. King and J.A.D. Wattis,   Asymptotic solutions of 
the Becker-D\"oring equations with size-dependent rate constants,
 J. Phys. A; Math. Gen., \textbf{35}, (2002) 1357--1380.

\bibitem{kp} M. Kreer, O. Penrose, Proof of dynamic scaling in
Smoluchowski's coagulation equations with constant kernels,
J. Stat. Phys., \textbf{74}, (1994) 389--407.

\bibitem{L99} Ph. Lauren\c{c}ot, Singular behaviour of 
finite approximations to the addition model,
Nonlinearity, \textbf{12}, (1999) 229--239.  

\bibitem{LM02} Ph. Lauren\c{c}ot, S. Mischler, From the Becker-D\"oring 
to the Lifshitz-Slyozov-Wagner equations,
J. Stat. Phys.,  \textbf{106}, (2002) 957--991.

\bibitem{LM} Ph. Lauren\c{c}ot, S. Mischler, On Coalescence 
Equations and Related Models, in P. Degond, L. Pareschi, G. Russo 
(Eds.), Modelling and Computational Methods for Kinetic 
Equations, Birkh\"auser, Boston, 2004, pp. 321--356. 

\bibitem{Ley03} F. Leyvraz, Scaling theory and exactly solved 
models in the kinetics of irreversible aggregation,
Phys. Rep., \textbf{383}, (2003) 95--212. 

\bibitem{lk} A.A. Lushnikov, M. Kulmala, Singular self-preserving regimes of
coagulation processes,
Phys. Rev. E, \textbf{65}, (2002) 041604.

\bibitem{MG} T. Matsoukas, E. Gulari, Monomer-addition growth with 
a slow initiation step: a growth model for silica particles from alkoxides,
J. Coll. Int. Sci., \textbf{132}, (1989) 13--21.

\bibitem{mp03-1} G. Menon, R.L. Pego, Approach to self-similarity 
in Smoluchowski's coagulation equations, 
Commun. Pure Appl. Math., \textbf{57}, (2004) 1197-1232.

\bibitem{mp03-2} G. Menon, R.L. Pego,  Dynamical scaling in 
Smoluchowski's coagulation equations: uniform convergence,
SIAM J. Math. Anal., \textbf{36}, (2005) 1629--1651.

\bibitem{N03} B. Niethammer, On the evolution of large clusters 
in the Becker-D\"oring model, J. Nonlinear Sci., \textbf{13}, (2003) 115-155.

\bibitem{p89} O. Penrose, Metastable states for the Becker-D\"oring
cluster equations, Commun. Math. Phys., \textbf{124}, (1989) 515--541.

\bibitem{pa} D.O. Pushkin, H. Aref, Self-similarity theory 
of stationary coagulation, Phys. Fluids, \textbf{14}, (2002) 694--703.

\bibitem{smol} M. von Smoluchowski, Versuch einer mathematischen Theorie
der Koagulationskinetik kolloider L\"osungen,
Z. Physik. Chem., \textbf{92}, (1917) 129--168.

\bibitem{w04} J.A.D. Wattis, Similarity solutions of a Becker-D\"oring system 
with time-dependent monomer input,
J. Phys. A; Math. Gen., \textbf{37}, (2004) 7823--7841.
 
\bibitem{wbc} J.A.D. Wattis, C.D. Bolton and P.V. Coveney, The 
Becker-D\"oring equations with exponentially size-dependent rate 
coefficients, J. Phys. A; Math. Gen., \textbf{37}, (2004) 2895--2912. 

\bibitem{wk} J.A.D. Wattis and J.R. King, Asymptotic solutions of the \BD\ equations, 
J. Phys. A; Math. Gen., \textbf{31}, (1998) 7169--7189. 

\bibitem{z92} Zhang Zhi-fen, Ding Tong-ren, Huang Wen-zao, Dong Zhen-xi, 
Qualitative Theory of Differential Equations,
Translations of Mathematical Monographs vol. 101, 
American Mathematical Society, Providence RI, 1992. 



\end{thebibliography}
\end{document}